\newif\ifAnon\Anonfalse
\newif\ifDraft\Draftfalse

\documentclass[conference]{IEEEtran}
\usepackage[hyphens]{url}
\PassOptionsToPackage{bookmarks,hidelinks}{hyperref}
\usepackage{hyperref}
\hypersetup{breaklinks=true}
\usepackage{tugraz_defaults}
\usepackage{xcolor,colortbl}
\usepackage{tcolorbox}
\usepackage{multirow}
\usepackage{listings}
\usepackage{parcolumns}

\graphicspath{{figures/}}

\usepackage[framemethod=tikz]{mdframed}
\mdfdefinestyle{mystyle}{%
  rightline=true,
  innerleftmargin=5,
  innerrightmargin=5,
  innertopmargin=0,
  outerlinewidth=0.5pt,
  topline=true,
  rightline=true,
  bottomline=true,
  roundcorner=1pt,
  skipabove=0.6\topskip,
  skipbelow=0.2\topskip
}

\mdfdefinestyle{tightstyle}{%
  rightline=true,
  innerleftmargin=2,
  innerrightmargin=2,
  innerbottommargin=0,
  innertopmargin=0,
  outerlinewidth=0.3pt,
  topline=true,
  rightline=true,
  bottomline=true,
  roundcorner=1pt,
  skipabove=0.6\topskip,
  skipbelow=0.2\topskip
}

\def\speciallstcolor{\begingroup\color{red}}
\definecolor{blue2}{rgb}{0.0, 0.0, 1.0}
\def\specialBlstcolor{\begingroup\color{blue2}}
\def\endspeciallstcolor{\endgroup}

\newcommand{\paragrabf}[1]{\noindent \textbf{#1}\ }

\usepgfplotslibrary{dateplot}
\usepgfplotslibrary{fillbetween}

\newcommand{\Attack}{Decomp+Time\xspace}
\newcommand{\Fuzzer}{Comprezzor\xspace}

\newcommand{\Crime}{CRIME\xspace}
\newcommand{\Time}{TIME\xspace}
\newcommand{\Breach}{BREACH\xspace}
\newcommand{\Heist}{HEIST\xspace}
\newcommand{\Deflate}{DEFLATE\xspace}
\newcommand{\Rupture}{Rupture}
\newcommand{\Zstd}{Zstd\xspace}

\begin{document}

\title{Practical Timing Side Channel Attacks on Memory Compression}

\date{}

\ifAnon
    \author{}
\else
    \author{\IEEEauthorblockN{Martin Schwarzl}
    \IEEEauthorblockA{Graz University of Technology\\
    martin.schwarzl@iaik.tugraz.at}
    \and
    \IEEEauthorblockN{Pietro Borrello}
    \IEEEauthorblockA{Sapienza University of Rome\\
    borrello@diag.uniroma1.it}
    \and
    \IEEEauthorblockN{Gururaj Saileshwar}
    \IEEEauthorblockA{Georgia Tech\\
    Graz University of Technology\\
    gururaj.s@gatech.edu}
    \and
    \IEEEauthorblockN{Hanna Müller}
    \IEEEauthorblockA{Graz University of Technology \\
    h.mueller@student.tugraz.at}
    \and
    \IEEEauthorblockN{Michael Schwarz}
    \IEEEauthorblockA{CISPA Helmholtz Center for Information Security \\
    michael.schwarz@cispa.saarland}
    \and
    \IEEEauthorblockN{Daniel Gruss}
    \IEEEauthorblockA{Graz University of Technology\\
    daniel.gruss@iaik.tugraz.at}}
\fi

% \IEEEoverridecommandlockouts
% \makeatletter\def\@IEEEpubidpullup{6.5\baselineskip}\makeatother
% \IEEEpubid{\parbox{\columnwidth}{
%     Network and Distributed Systems Security (NDSS) Symposium 2021\\
%     24-26 February 2021, San Diego, CA, USA\\
%     ISBN 1-891562-61-4\\
%     https://dx.doi.org/10.14722/ndss.2021.23xxx\\
%     www.ndss-symposium.org
% }
% \hspace{\columnsep}\makebox[\columnwidth]{}}

\maketitle

\begin{abstract}
Compression algorithms are widely used as they save memory without losing data.
However, elimination of redundant symbols and sequences in data leads to a compression side channel.
So far, compression attacks have only focused on the compression-ratio side channel, \ie the size of compressed data, and largely targeted HTTP traffic and website content. 

In this paper, we present the first memory compression attacks exploiting timing side channels in compression algorithms, targeting a broad set of applications using compression.
Our work systematically analyzes different compression algorithms and demonstrates timing leakage in each.
We present \Fuzzer, an evolutionary fuzzer which finds memory layouts that lead to amplified latency differences for decompression and therefore enable remote attacks. 
We demonstrate a remote covert channel exploiting small local timing differences transmitting on average \SI{643.25}{\bit/\hour} over \SIx{14} hops over the internet.
We also demonstrate memory compression attacks that can leak secrets bytewise as well as in dictionary attacks in three different case studies.
First, we show that an attacker can disclose secrets co-located and compressed with attacker data in PHP applications using Memcached.
Second, we present an attack that leaks database records from PostgreSQL, managed by a Python-Flask application, over the internet.
Third, we demonstrate an attack that leaks secrets from transparently compressed pages with ZRAM, the memory compression module in Linux.
We conclude that memory-compression attacks are a practical threat.
\end{abstract}

\section{Introduction}

Data compression plays a vital role for performance and efficiency.
For example, compression leads to smaller data footprints, allowing more data to be stored in memory.
Memory accesses are typically faster than retrieving data from slower mediums such as hard disks or solid-state disks.
As a result, both Microsoft~\cite{Yosifovich2017} and Apple~\cite{AppleMemoryCompression} rely on memory compression in their operating systems (OSs) to better utilize memory.
Similarly, memory compression is also used in databases~\cite{PostgreSQL2021TOASTCompression} or key-value stores~\cite{PHP2021Memcached}.
Compression can also increase performance when storing or transferring data on slow mediums. 
Hence, data compression is also widely used for HTTP traffic~\cite{Mozilla2021Compression,Fielding1999HTTP} and file-system compression~\cite{Btrfs2021Compression}.

While compression has many advantages, it is problematic in scenarios where sensitive data is compressed~\cite{Rizzo2012Crime,Gluck2013Breach,Beery2013TIME,VanHoef2016Heist,VanGoethem2016Request,Karakostas2016Practical}. 
Attacks such as \Crime~\cite{Rizzo2012Crime}, \Breach~\cite{Gluck2013Breach}, \Time~\cite{Beery2013TIME}, or \Heist~\cite{VanHoef2016Heist}, exploit the combination of encryption and compression in TLS-encrypted HTTP traffic. 
Karaskostas~\etal\cite{Karakostas2016Practical} extended \Breach to commonly used ciphers, such as AES.
These attacks demonstrated that an attacker could learn secrets when they are compressed together with attacker-controlled input.
However, all these attacks focused on web traffic and only exploited differences in the compressed size of data. 
The size of compressed data is either accessed directly~\cite{Rizzo2012Crime}, \eg in a man-in-the-middle (MITM) attack or indirectly by observing the impact of the compressed size on the transmission time~\cite{Beery2013TIME}. 

Although attacks exploiting the compression ratio were first described by Kelsey~\etal\cite{Kelsey2002Compression} in 2002, most attacks thus far have focused on compressed web traffic. Surprisingly, security implications of compression in other settings, such as virtual memory and file systems, have not been studied thoroughly.
Also, compression fundamentally provides a reduction in size of data at the expense of additional time required for compression and decompression.
However, so far, only the \emph{result} of the compression, \ie the compressed size, has been exploited to leak data but not the time consumed by the \emph{process} of compression or decompression itself. 

In this paper, we demonstrate that the \emph{decompression time} also leaks information about the compressed data. 
Instead of focussing on the final compressed data, we measure the time it takes to decompress data. 
Hence, we do not need to observe the transport of the compressed data.
Decompression timing can be measured in several scenarios, \eg a simple read access to the data can suffice for compressed virtual memory or file systems. 
Moreover, decompression timings can be observed remotely, even when the compressed data never leaves the victim system.

We systematically analyzed six compression algorithms, including widely-used algorithms such as \Deflate (in zlib), PGLZ (in PostgreSQL), and zstd (by Facebook).
We observe that the decompression time not only correlates with the entropy of the uncompressed data, but also with various other aspects, such as the relative position or alignment of compressible data.
In general, these timing differences arise due to the design of the compression algorithm and its implementation. 

To explore these parameters in an automated fashion, we introduce \Fuzzer, an evolutional fuzzer. 
In line with previous attacks on compression, \Fuzzer assumes that parts of the compressed data are controlled by an attacker. 
Based on this assumption and a target secret, \Fuzzer searches for a memory layout that maximizes the timing differences for decompression timing attacks.
As a result, \Fuzzer discovers a memory layout for compression algorithms which leads to decompression timing attacks with high-latency differences up to three orders of magnitude higher than manually discovered layouts. 

Based on the results of \Fuzzer, we present three case studies demonstrating that these timing differences can be exploited in realistic scenarios to leak sensitive data.
We demonstrate that these timing differences can be exploited even in remote attacks on an in-memory database system without executing code on the victim machine and without observing the victim's network traffic. 
Hence, our case studies show that compressing sensitive data poses a security risk in any scenario using compression and not just for web traffic. 

In the first case study, we build a remote covert channel which works across $14$ hops on the internet, abusing the memory compression of Memcached, an in-memory object caching system. 
By measuring the execution time of a public PHP API of an application using Memcached, we can covertly transfer data between two computers at an average transmission rate of \SI{643.25}{\bit/\hour} with an error rate of only \SI{0.93}{\percent}. 
In a similar setup, we also demonstrate the capability of leaking a secret bytewise co-located with data compressed by a PHP application into Memcached, in \SI{31.95}{\min} ($n=20,\sigma=8.48\%$) over the internet.
Using a dictionary with $100$ guesses, we leak a \SI{6}{\byte}-secret in \SI{14.99}{\min}.

In the second case study, we exploit the transparent database compression of PostgreSQL.
Our exploit leaks database records from an inaccessible database table of a remote web app.
We show that as long as an attacker can co-locate data with a secret, it is possible for the attacker to influence and observe decompression times when accessing the data, thus, leaking secret data.
Such a scenario might arise if structured data, \ie JSON documents, are stored with attack-controlled fields into a single cell.
Using an attack that leaks a secret bytewise, we leak a \SI{6}{\byte}-secret in \SI{7.85}{\min}.
Our dictionary attack runs in \SI{8.28}{\min} on average over the internet for $100$ guesses.

In the third case study, we exploit ZRAM, the memory compression module in Linux. 
We demonstrate that even if the application itself or the file system does not compress data, the OS's memory compression can transparently introduce timing side channels. 
We demonstrate that a dictionary attack with 100 guesses on ZRAM decompression can leak a \SI{6}{\byte}-secret co-located with attacker data in a page within \SI{2.25}{\min} on average.

Our fuzzer shows that most if not all compression algorithms are susceptible to timing side channels when observing the decompression of data. 
With our case studies, we demonstrate that data leakage is not only a concern for data in transit but also for data at rest. 
With this work, we seek to highlight the importance of evaluating the trade-off between compressing data, and leaking side-channel information about the compressed data, for any system adopting compression. 

\subheading{Contributions.} The main contributions of this work are:
\begin{compactenum}
\item We present a systematic analysis of timing leakage for several lossless data-compression algorithms. 
\item We demonstrate an evolutional fuzzer to automatically search for memory layouts causing large timing differences for timing side channels on memory compression.
\item We show a remote covert channel exploiting the in-memory compression of PHP-Memcached.
\item We demonstrate that compression-based timing side channels can be introduced by compression in applications, databases, or the system’s memory compression.
\item We leak secrets from Memcached, PostgreSQL, and ZRAM within minutes. 
\end{compactenum}

\subheading{Outline.}
In \cref{sec:background}, we provide background.
In \cref{sec:attack}, we present attack model and building blocks.
In \cref{sec:analysis}, we systematically analyze compression algorithms.
In \cref{sec:fuzzing}, we present \Fuzzer.
In \cref{sec:casestudies}, we demonstrate local and remote attacks exploiting decompression timing.
We discuss mitigations in \cref{sec:discussion} and conclude in \cref{sec:conclusion}.

We responsibly disclosed our findings to the developers and will open-source our fuzzing tool and attacks on Github.%\footnote{\url{https://github.com/anonymized-for-submission}}.

\section{Background and Related Work}\label{sec:background}

In this section, we provide background on data compression algorithms, prior side-channel attacks on data compression, and the use of fuzzing to discover side channels.

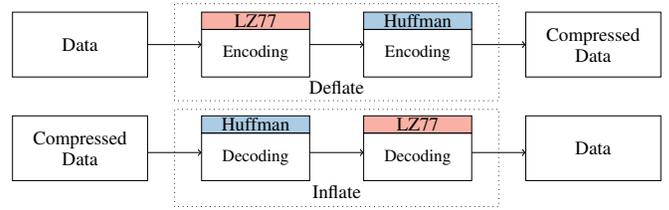
\begin{figure}[t]
  \centering
  \resizebox{\hsize}{!}{
      \begin{tikzpicture}[yscale=0.8]

\begin{scope}[shift={(0,-0.8)}]
\draw[dotted] (-0.5,1.45) rectangle +(6,2.25);
\draw[->] (-1,2.75) -- (0,2.75);
\draw[->] (2,2.75) -- (3,2.75);
\draw[->] (5,2.75) -- (6,2.75);

\draw (-3.5,2) rectangle +(2.5,1.5) node[pos=.5,yshift=-0.0cm] {\parbox{2.25cm}{\centering Data}};

\draw[fill=red!40] (0,3.1) rectangle +(2,0.4) node[pos=.5] {LZ77};
\draw (0,2) rectangle +(2,1.1) node[pos=.5] {\parbox{2cm}{\centering \small Encoding}};

\draw[fill=blue!40] (3,3.1) rectangle +(2,0.4) node[pos=.5] {Huffman};
\draw (3,2) rectangle +(2,1.1) node[pos=.5] {\parbox{2cm}{\centering \small Encoding}};

\draw (6,2) rectangle +(2.5,1.5) node[pos=.5,yshift=-0.0cm] {\parbox{2.25cm}{\centering Compressed Data}};

\node at (2.5,1.75) {Deflate};

\end{scope}

\draw[dotted] (-0.5,-1.8) rectangle +(6,2.25);

\draw[->] (-1,-.55) -- (0,-.55);
\draw[->] (2,-.55) -- (3,-.55);
\draw[->] (5,-.55) -- (6,-.55);

\draw (-3.5,-1.2) rectangle +(2.5,1.5) node[pos=.5,yshift=-0.0cm] {\parbox{2.25cm}{\centering Compressed Data}};

\draw[fill=blue!40] (0,-0.1) rectangle +(2,0.4) node[pos=.5] {Huffman};
\draw (0,-1.2) rectangle +(2,1.1) node[pos=.5] {\parbox{2cm}{\centering \small Decoding}};

\draw[fill=red!40] (3,-0.1) rectangle +(2,0.4) node[pos=.5] {LZ77};
\draw (3,-1.2) rectangle +(2,1.1) node[pos=.5] {\parbox{2cm}{\centering \small Decoding}};

\draw (6,-1.2) rectangle +(2.5,1.5) node[pos=.5,yshift=-0.0cm] {\parbox{2.25cm}{\centering Data}};

\node at (2.5,-1.45) {Inflate};

\end{tikzpicture}
  }
  \caption{\Deflate algorithm has two parts: LZ77 part to compress sequences and a Huffman part to compress symbols.}
  \label{fig:deflate}
\end{figure}

\subsection{Data Compression Algorithms}
Lossless compression reduces the size of data without losing information.
One of the most popular algorithms is the \Deflate compression algorithm, which is used in gzip (zlib).
The \Deflate compression algorithm~\cite{Deutsch1996Deflate} consists of two main parts, LZ77 and Huffman encoding and decoding, as shown in \cref{fig:deflate}. 
The \texttt{Lempel-Ziv} (LZ77) part scans for the longest repeating sequence within a sliding window and replaces repeated sequences with a reference to the first occurrence~\cite{Euccas2021UnderstandingZlib}. 
This reference stores distance and length of the occurrence.
The \texttt{Huffman}-coding part tries to reduce the redundancy of symbols.
When compressing data, \Deflate first performs LZ77 encoding and Huffman encoding~\cite{Euccas2021UnderstandingZlib}.
When decompressing data (inflate), they are performed in reverse order.
The algorithm provides different compression levels to optimize for compression speed or compression ratio.
The smallest possible sequence has a length of \SI{3}{\byte}\cite{Deutsch1996Deflate}.

Other algorithms provide different design points for compressibility and speed.
\Zstd, designed by Facebook~\cite{Collett2021Zstd} for modern CPUs, improves both compression ratio and speed, and is used for compression in file systems (\eg btrfs, squashfs) and databases (\eg AWS Redshift, RocksDB).
LZ4 and LZO are other algorithms used in file systems, which are optimized for compression and decompression speed. 
LZ4, in particular, gains its performance by using a sequence compression stage (LZ77) without the symbol encoding stage (Huffman) like in \Deflate.
FastLZ, similar to LZ4, is a fast compression algorithm implementing LZ77.
PGLZ is a fast LZ-family compression algorithm used in PostgreSQL for varying-length data in the database~\cite{PostgreSQL2021TOASTCompression}.

\subsection{Prior Data Compression Attacks}\label{sec:previous_attacks}
In 2002, Kelsey~\cite{Kelsey2002Compression} first showed that any compression algorithm is susceptible to information leakage based on the compression-ratio side channel.
Duong and Rizzo~\cite{Rizzo2012Crime} applied this idea to steal web cookies with the \Crime attack by exploiting TLS compression.
In the \Crime attack, the attacker adds additional sequences in the HTTP request, which act as guesses for possible cookies values, and observes the request packet length, \ie the compression ratio of the HTTP header injected by the browser.
If the guess is correct, the LZ77-part in gzip compresses the sequence, making the compression ratio higher, thus allowing the secret to be discovered. 
To perform \Crime, the attacker needs to be able to spy on the packet length, and the secret needs to have a known prefix such as \texttt{sessionid=} or \texttt{cookie=}.
To mitigate \Crime, TLS-level compression was disabled for requests~\cite{Beery2013TIME,Gluck2013Breach}. 

The \Breach attack~\cite{Gluck2013Breach} revived the \Crime attack by attacking HTTP responses instead of requests and leaking secrets in the HTTP responses such as cross-site-request-forgery tokens.
The \Time attack~\cite{Beery2013TIME} uses the time of a response as a proxy for the compression ratio, as it can be measured even via JavaScript. 
To reliably amplify the signal, the attacker chooses the size of the payload such that additional bytes, due to changes in compressibility, cross a boundary and cause significantly higher delays in the round-trip time  (RTT).
\Time exploits the compression ratio to amplify timing differences via TCP windows and does not exploit timing differences in the underlying compression algorithm itself. 

A malicious site can perform a \Time attack on their visitors to break same-origin policy or leak data from sites that reflect user input in the response, such as a search engine~\cite{Beery2013TIME}.
Vanhoef and Van Goethem~\cite{VanHoef2016Heist} showed with \Heist that HTTP/2 features can also be used to determine the size of cross-origin responses and to exploit \Breach using the information.
Van Goethem~\etal~\cite{VanGoethem2016Request} similarly showed that compression can be exploited to determine the exact size of any resource in browsers. 
Karaskostas and Zindros\cite{Karakostas2016Practical} presented \Rupture, extending \Breach attacks to web apps using block ciphers such as AES.

Tsai~\etal\cite{Tsai2020SafeCracker} demonstrated cache timing attacks on compressed caches, which can leak a secret key in under \SI{10}{\milli\second}.

\textbf{Common Theme.} All of the prior attacks primarily exploit the compression-ratio side channel. However, the time taken by the underlying compression algorithm for compression or decompression is not analyzed or exploited as side channels. Additionally, these attacks largely target the HTTP traffic and website content, and do not focus on the broader applications of compression such as memory-compression, databases, file systems, and others, that we target in this paper.

\subsection{Fuzzing to Discover Side Channels}
Historically, fuzzing has been used to discover bugs in applications~\cite{Zalewski2021AFL, Fioraldi2021AFLplusplus}.
Typically, it involves feedback based on novelty search, executing inputs, and collecting ones that cover new program paths in the hope of triggering bugs. 
Other fuzzing proposals use genetic algorithms to direct input generation towards interesting paths~\cite{Rawat2017, She2019Neuzz}.
Recently fuzzing has also been used to discover side channels both in software and in the microarchitecture~\cite{Weber2021osiris, Gras2020, Moghimi2020medusa, Fogh2016shotgun}.
\texttt{ct-fuzz}~\cite{Shaobo2020ctfuzz} used fuzzing to discover timing side channels in cryptographic implementations.
Nilizadeh~\etal\cite{Nilizadeh2019Diffuzz} used differential fuzzing to detect compression-ratio side channels that enable the \Crime attack. 
In this work, we build on these and use fuzzing to discover timing channels in compression algorithms.

\section{High-level Overview}\label{sec:attack}
In this section, we discuss the high-level overview of memory compression attacks and the attack model.
\subsection{Attack Model \& Attack Overview}\label{sec:attack_overview}
Most prior attacks discussed in~\Cref{sec:previous_attacks} focused on the compression ratio side channel.
Even the \Time attack and its variants by Vanhoef and Van Goethem~\etal\cite{VanHoef2016Heist,VanGoethem2016Request} only exploited timing differences due to the TCP protocol.
None of these exploited or analyzed timing differences due to the compression algorithm itself, which is the focus of our attack.

We assume that the attacker can co-locate arbitrary data with secret data.
This co-location can be given, \eg via a memory storage API like Memcached or a shared database between the attacker and the victim.
Once the attacker data is compressed with the secret, the attacker only needs to measure the latency of a subsequent access to its data.
Furthermore, we assume no software vulnerabilities \ie memory corruption vulnerabilities in the application it self.

\begin{figure}[t]
    \centering
    \resizebox{\hsize}{!}{
        \input{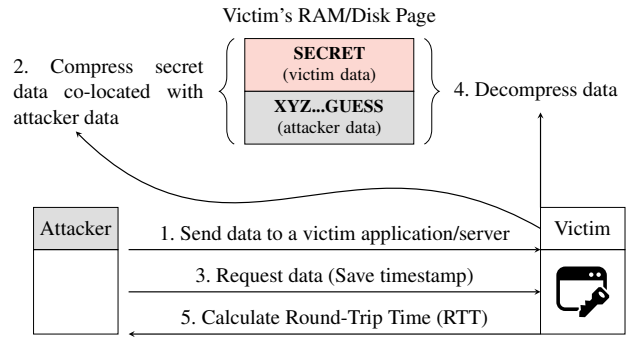}
    }
    \caption{Overview of a memory compression attack exploiting a timing side channel.}
    \label{fig:attack_overview}
\end{figure}

\cref{fig:attack_overview} illustrates an overview of a memory compression attack in five steps.
The victim application can be a web server with a database or software cache, or a filesystem that compresses stored files.
\textbf{First}, the attacker sends its data to be stored to the victim's application.
\textbf{Second}, the victim application compresses the attacker-controlled data, together with some co-located secret data, and stores the compressed data.
The attacker-controlled data contains a partial guess of the co-located victim's data \texttt{SECRET} or, in the case where a prefix is known, \texttt{prefix=SECRET}.
The guess can be performed bytewise to reduce the guessing entropy.
If the partial guess is correct, \ie \texttt{SECR}, the compressed data not only has a higher compression ratio, but it also influences the decompression time.
\textbf{Third}, after the compression happened, the attacker requests the content of the stored data again and takes a timestamp.
\textbf{Fourth}, the victim application decompresses the input data and responds with the requested data.
\textbf{Fifth}, the attacker takes another timestamp when the application responds and computes the RTT as the difference between the two timestamps.
Based on the RTT, which depends on the decompression latency of the algorithm, the attacker infers whether the guess was correct or not and thus infers the secret data.
Thus, the attack relies on the \textit{timing differences} of the compression algorithm itself, which we characterize next.

\section{Systematic Study: Compression Algorithms}\label{sec:analysis}

In this section, we provide a systematic analysis of timing leakage in compression algorithms. 
We choose six popular compression algorithms (zlib, zstd, LZ4, LZO, PGLZ, and FastLZ), and evaluate the compression and decompression times based on the entropy of the input data.
Zlib (which implements the DEFLATE algorithm) is the most popular as a standard for compressing files and is used in gzip.
Zstd is Facebook's alternative to Zlib.
PGLZ is used in the popular relational database management system PostgreSQL.
LZ4, FastLZ, and LZO were built to increase compression speeds. 
For each of the algorithms, we observe timing differences in the range of hundreds to thousands of nanoseconds based on the content of the input data (\SI{4}{\kilo\byte}-page).

\subsection{Experimental Setup}\label{sec:setup}
We conducted the experiments on an Intel i7-6700K (Ubuntu 20.04, kernel 5.4.0) with a fixed frequency of \SI{4}{\giga\hertz}.
We evaluate the latency of each compression algorithm with three different input values, each \SI{4}{\kilo\byte} in size. 
The first input is the same byte repeated 4096 times, which should be \textit{fully compressible}.
The second input is \textit{partly compressible} and a hybrid of two other inputs: half random bytes and half compressible repeated bytes.
The third input  consists of random bytes which are theoretically \textit{incompressible}.
With these, we show that compression algorithms can have different timings dependent on the compressibility of the input.

\subsection{Timing Differences for Different Inputs} \label{sec:TimingVariationsDecompression}
For each algorithm and input, we measure the decompression and compression time over \SIx{100000} repetitions and compute the mean values and standard deviations.

\begin{table}[t]
\setlength{\aboverulesep}{0pt}
\setlength{\belowrulesep}{0pt}
\caption{Different compression algorithms yield distinguishable timing differences when decompressing content with a different entropy ($n=100000$).}
\label{tab:evaluated_algorithms_decomp}
\begin{center}
\vspace{-0.2cm}
\resizebox{\hsize}{!}{
  \begin{tabular}{|l|r|r|r|}
      \hline
      \multirow{2}{*}{Algorithm}                    & Fully~\qquad\qquad & Partially~\qquad\quad & \multirow{2}{*}{Incompressible (ns)} \\
      & Compressible (ns) & Compressible (ns) & \\
      \hline \addlinespace[0.2cm]
      FastLZ & \SIx{7257.88} ($\pm0.23\%$) & \SIx{4264.56} ($\pm2.27\%$) & \SIx{1155.57} ($\pm0.92\%$) \\
      LZ4  & \SIx{605.79} ($\pm1.02\%$) & \SIx{218.68} ($\pm1.76\%$) & \SIx{107.90} ($\pm2.49\%$) \\
      LZO  & \SIx{2115.65} ($\pm2.05\%$) & \SIx{1220.07} ($\pm3.64\%$) & \SIx{309.44} ($\pm6.27\%$) \\
      PGLZ & \SIx{813.75} ($\pm0.71\%$) & \SIx{5340.47} ($\pm0.38\%$) & - \\
      zlib & \SIx{7016.02} ($\pm0.33\%$) & \SIx{13212.53} ($\pm0.35\%$) & \SIx{1640.09} ($\pm1.51\%$) \\
      zstd & \SIx{941.05} ($\pm0.94\%$) & \SIx{772.55} ($\pm0.77\%$) & \SIx{370.59} ($\pm2.87\%$) \\\addlinespace[0.1cm]
      \hline
  \end{tabular}
}
  \end{center}
\end{table}
  
\paragrabf{Decompression.}
\Cref{tab:evaluated_algorithms_decomp} lists the decompression latencies for all evaluated compression algorithms.
Depending on the entropy of the input data, there is considerable variation in the decompression time. 
All algorithms incur a higher latency for decompressing a fully compressible page compared to an incompressible page, leading to a timing difference of few hundred to few thousand nanoseconds for different algorithms. 
This is because, for incompressible data, algorithms can augment the raw data with additional metadata to identify such cases and perform simple memory copy operations to ``decompress'' the data, as is the case for zlib where the decompression for an incompressible page is \SI{5375.93}{\nano\second} faster than a fully compressible page.
For decompression of partially compressible pages, some algorithms (FastLZ, LZ4, LZO, zstd) have lower latency than fully compressible pages, while others (zlib and PGLZ) have a higher latency than fully compressible pages. 
This shows the existence of even  algorithm-specific variations in timings. 
PGLZ does not create compressible memory in the case of an incompressible input, and hence we do not measure its latency for this input.

\paragrabf{Compression.} 
For compression, we observed a trend in the other direction (\Cref{tab:evaluated_algorithms_comp} in~\cref{sec:appendix-timing_diff_compression} lists compression latencies for different algorithms).
For a fully compressible page, there are also latencies between the three different inputs, which are clearly distinguishable in the order of multiple hundreds to thousands of nanoseconds. 
Thus, timing side channels from compression might also be used to exploit compression of attacker-controlled memory co-located with secret memory.
However, attacks using the compression side channel might be harder to perform in practice as the compression of data might be performed in a separate task (in the background), and the latency might not be easily observable for a user. 
Hence, our work focuses on attacks exploiting the decompression timing side channel.

\paragrabf{Handling of Corner Cases.}
For incompressible pages, the ``compressed'' data can be larger than the original size with the additional compression metadata. 
Additionally, it is slower to access after compression than raw uncompressed data. 
Hence, this corner-case with incompressible data may be handled in an implementation-specific manner, which can itself lead to additional side channels.
For example, a threshold for the compression ratio can decide when a page is stored in a raw format or in a compressed state, like in Memcached-PHP~\cite{PHP2021Memcached}.
Alternatively, PGLZ, the algorithm used in PostgreSQL database, which computes the maximum acceptable output size for input by checking the input size and the strategy compression rate, could simply fail to compress inputs in such corner cases.

In~\Cref{sec:casestudies}, we show how real-world applications like Memcached, PostgreSQL, and ZRAM deal with such corner cases and demonstrate attacks on each of them.

\subsection{Leaking secrets via timing side channel}\label{sec:dictionary-attack}
Thus far, we analyzed timing differences for decompressing different inputs, which in itself is not a security issue.
In this section, we demonstrate \Attack to leak secrets from compressed pages using these timing differences.
We focus on sequence compression, \ie LZ77 in \Deflate.

\subsubsection{Building Blocks for \Attack}
\Attack consists of three building blocks: \textit{sequence compression} that we use to modulate the compressibility of an input, \textit{co-location} of attacker data and secrets, and \textit{timing variation for decompression} depending on the change in compressibility of the input.

\paragrabf{Sequence compression:}
Sequence compression \ie LZ77 tries to reduce the redundancy of repeated sequences in an input by replacing each occurrence with a pointer to the first occurrence. 
This results in a higher compression ratio if redundant sequences are present in the input and a lower ratio if no such sequences are present.
This compressibility side channel can leak information about the compressed data.

\paragrabf{Co-location of attacker data and secrets:}
If the attacker can control a part of data that is compressed with a secret, as described in \Cref{fig:attack_overview}, then the attacker can place a \textit{guess} about the secret and place it co-located with the secret to exploit sequence compression. 
If the compression ratio increases, the attacker can infer if the guess matches the secret or not. 
While the \Crime attack~\cite{Rizzo2012Crime} previously used a similar set up and observed the compressed size of HTTP requests to steal secrets like HTTP cookies, we target a more general attack and do not require observability of compressed sizes.

\paragrabf{Timing Variation in Decompression:}
We infer the change in compressibility via its influence on decompression timing. 
We observe that even sequence compression can cause variation in the decompression timing based on compressibility of inputs (for all algorithms in \cref{sec:TimingVariationsDecompression}). 
If the sequence compression reduces redundant symbols in the input and increases the compression ratio, we observe faster decompression due to fewer symbols.
Otherwise, with a lower compression ratio and more number of symbols, decompression is slower.
Hence, the attacker can infer the compressibility changes for different guesses by observing differences in decompression time due to sequence compression. 
For a correct guess, the guess and the secret are compressed together and the decompression time is faster due to fewer symbols.
Otherwise, it is slower for incorrect guesses with more symbols.

\subsubsection{Launching the \Attack}
Using the building blocks described above, we setup the attack with an artificial victim program that has a \SI{6}{\byte}-secret string (\texttt{SECRET}) embedded in a \SI{4}{\kilo\byte}-page. 
The page also contains attacker-controlled data that is compressed together with the secret, like the scenario shown in \Cref{fig:attack_overview}.
The attacker can update its own data in place, allowing it to make multiple guesses. 
The attacker can also read this data, which triggers a decompression of the page and allows the attacker to measure the decompression time. 
A correct guess that matches with the secret results in faster decompression. 

We perform the attack on the zlib library (1.2.11) and use \SIx{8} different guesses, including the correct guess. 
For each guess, a single string is placed \SI{512}{\byte} away from the secret value; other data in the page is initialized with dummy values (repeated number sequence from 0 to 16).
To measure the execution time, we use the \texttt{rdtsc} instruction.

\begin{figure}[t]
  \centering
  \resizebox{\hsize}{!}{
      \begin{tikzpicture}
\begin{axis}[
style={font=\footnotesize},
xlabel={Guess},
ylabel={Timing [ns]},
y label style={align=center,text width=2cm},
width=\hsize,
height=3cm,
xtick={0,40,80,120,160,200,240,280},
xticklabels={FOOBAR,\textbf{SECRET},PYTHON,123456,COOKIE,SOMEDA,ADMIN1,NOPENO},
x tick label style={rotate=45,anchor=east}
]

\addplot+[mark=*,draw=none] table[x=Index,y=Time,col sep=semicolon] {data/basic-dict.csv};
\addplot+[mark=*,only marks,error bars/.cd,y dir=both,y explicit]
        table[x=Index,y=Time,col sep=semicolon,
        restrict expr to domain={\thisrow{Index}}{40:40}
        ]{data/basic-dict.csv};
\draw[draw=none,fill=red] (axis cs:-30,3950) rectangle (axis cs:310,3954);

\end{axis}
\end{tikzpicture}}
  \caption{Decompression time with \Attack for a dictionary attack. There is a clear threshold (line) separating the correct from wrong secrets.}
  \label{fig:attack_simple}
\end{figure}
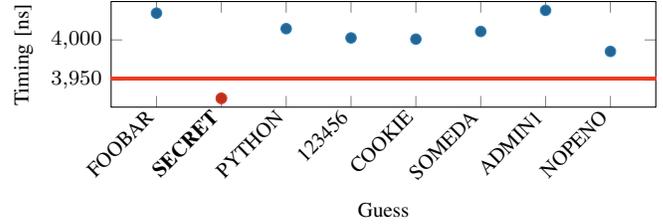

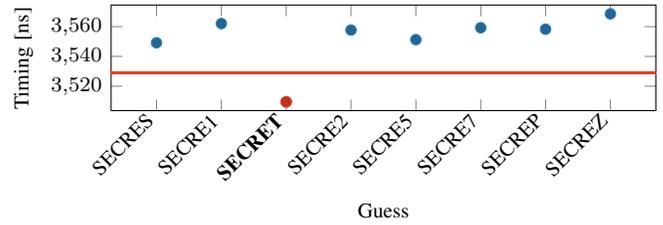
\begin{figure}[t]
  \centering
  \resizebox{\hsize}{!}{
      \begin{tikzpicture}
\begin{axis}[
style={font=\footnotesize},
xlabel={Guess},
ylabel={Timing [ns]},
y label style={align=center,text width=2cm},
width=\hsize,
height=3cm,
xtick={0,40,80,120,160,200,240,280},
xticklabels={SECRES,SECRE1,\textbf{SECRET},SECRE2,SECRE5,SECRE7,SECREP,SECREZ,},
x tick label style={rotate=45,anchor=east}
]

\addplot+[mark=*,draw=none] table[x=Index,y=Time,col sep=semicolon] {data/basic-bytewise.csv};
\addplot+[mark=*,only marks,error bars/.cd,y dir=both,y explicit]
        table[x=Index,y=Time,col sep=semicolon,
        restrict expr to domain={\thisrow{Index}}{80:80}
        ]{data/basic-bytewise.csv};
\draw[draw=none,fill=red] (axis cs:-30,3530) rectangle (axis cs:310,3528);
\end{axis}
\end{tikzpicture}}
  \caption{Leaking the secret sixth offset character-wise. Still, the correct guess is distinguishable from the wrong guesses. A clear threshold can be used to separate the correct guess from the incorrect guesses.}
  \label{fig:attack_simple_bytewise}
\end{figure}

\paragrabf{Evaluation.}
Our evaluation was performed on an Intel i7-6700K (Ubuntu 20.04, kernel 5.4.0) with a fixed frequency of \SI{4}{\giga\hertz}. 
To get stable results, we repeat the decompression step with each guess \SIx{10000} times and repeat the entire attack \SIx{100} times.
For each guess, we take the minimum timing difference per guess and choose the global minimum timing difference to determine the correct guess.
\Cref{fig:attack_simple} illustrates the minimum decompression times. 
With zlib, we observe that the correct guess is faster on average by \SI{71.5}{\nano\second} ($n=100,\sigma=27.91\%$) compared to the second-fastest guess.
Our attack correctly guessed the secret in all 100 repetitions of the attack.

While we used a \SI{6}{\byte} secret, our experiment also works for smaller secrets down to a length of \SI{4}{\byte}.
We also observe the attack is faster if the secret has a known prefix with a length of \SI{3}{\byte} or more.

\paragrabf{Bytewise leakage.}\
If the attacker manages to guess or know the first three bytes of the secret, the subsequent bytes can even be leaked bytewise using our attack.
Both \Crime and \Breach, assume a known prefix such as \texttt{cookie=}.
Similar to \Crime and \Breach~\cite{Rizzo2012Crime,Gluck2013Breach,Karakostas2016Practical}, we try to perform a bytewise attack by modifying our simple layout.
We use the first $5$ characters of $SECRET$ as a prefix \texttt{"SECRE"} and guess the last byte with $7$ different guesses.
On average the latency is \SI{28.37}{\nano\second} ($n=100,\sigma=65.78\%$), between the secret and second fastest guess.
\Cref{fig:attack_simple_bytewise} illustrates the minimum decompression times for the different guesses. 
However, we observe an error rate of \SI{8}{\percent} for this experiment, which might be caused by the Huffmann-decoding part in \Deflate.

While techniques like the Two-Tries method~\cite{Rizzo2012Crime,Gluck2013Breach,Karakostas2016Practical} have been proposed to overcome the effects of Huffman-coding in \Deflate to improve the fidelity of byte-wise attacks exploiting compression ratio, we seek to explore whether bytewise leakage can be reliably performed via the timing only by amplying the timing differences.

\subsubsection{Challenge Of Amplifying Timing}
While the decompression timing side channels can be directly used in attacks as demonstrated, the timing differences are still quite small for practical exploits on real-world applications. 
For example, the timing differences we observed for the correct guess are in tens of nanoseconds, while most practical use-cases of compression such as in a Memcached server accessed over the network, or PostgreSQL database accessed from a filesystem could have access latencies of milliseconds. 

\paragrabf{Amplification.}\label{sec:amplification}
To enable memory compression attacks via the network or on data stored in file systems, we need to amplify the timing difference between correct and incorrect guesses.
However, it is impractical to manually identify inputs that could amplify the timing differences, as each compression algorithm has a different implementation that is often highly optimized.  
Moreover, various input parameters could influence the timing of decompression, such as frequency of sequences, alignments of the secret and attack-controlled data, size of the input, entropy of the input, and different compression levels provided by algorithms.
As an additional factor, the underlying hardware microarchitecture might also have undesirable influences.
Therefore, to automate this process in a systematic manner, we develop an evolutionary fuzzer, \Fuzzer, to find input corner cases that amplify the timing difference between correct and incorrect guesses for different compression algorithms.

\section{Evolutionary Compression-Time Fuzzer}\label{sec:fuzzing}
Compression algorithms are highly optimized, complex and even though most of them are open-source, their internals are only partially documented.
Hence, we introduce \Fuzzer, an evolutionary fuzzer to discover attacker-controlled inputs for compression algorithms that maximize  differences in decompression times for certain guesses enabling both bytewise leakage and dictionary attacks. 

\Fuzzer empowers genetic algorithms to amplify decompression side channels.
It treats the decompression process of a compression algorithm as an opaque box and mutates inputs to the compression while trying to maximize the output, \ie timing differences for decompression with different guesses.
The mutation process in \Fuzzer focuses on not only the entropy of the data, but also the memory layout and alignment that end up triggering optimizations and slow paths in the implementation that are not easily detectable manually.

While previous approaches used fuzzing to detect timing side channels~\cite{Nilizadeh2019Diffuzz, Shaobo2020ctfuzz}, \Fuzzer can dramatically amplify timing differences by being specialized for compression algorithms by varying parameters like the input size, layout, and entropy that affect the decompression time. The inputs discovered by \Fuzzer can amplify timing differences to such an extent that they are even observable remotely. 

\subsection{Design of \Fuzzer} 
In this section, we describe the key components of our fuzzer: Input Generation, Fitness Function, Input Mutation and Input Evolution.

\paragrabf{Input Generation.}
\Fuzzer generates memory layouts for \Attack by maximizing the timing differences on decompression of the correct guess compared to incorrect ones.
\Fuzzer creates layouts with sizes in the range of \SIrange{1}{64}{\kilo\byte}.
It uses a helper program that takes the memory layout configuration as input, builds the requested memory layout for each guess, compresses them using the target compression algorithm, and reports the observed timing differences in the decompression times among the guesses.
A memory layout configuration is composed of a file to start from, the offset of the secret in the file, the offset of guesses, how often the guesses are repeated in the layout, the compression level (\ie between 1 and 9 for zlib), and a modulus for entropy reduction that reduces the range of the random values.
The fuzzer can be used in cases where a prefix is known and unknown.

\paragrabf{Fitness Function.}
The evolutionary algorithm of \Fuzzer starts from a random population of candidate layouts (samples) and takes as feedback the difference in time between decompression of the generated memory containing the correct guess and the incorrect ones. 
\Fuzzer uses the timing difference between the correct guess and the second-fastest guess as the fitness score for a candidate.

The fitness function is evaluated using a helper program performing an attack on the same setup as in~\Cref{sec:setup}.
The program performs \SIx{100} iterations per guess and reports the minimum decompression time per guess to reduce the impact of noise.
This minimum decompression time is the output of the fitness function for \Fuzzer. 

\paragrabf{Input Mutation.}
\Fuzzer is able to amplify timing differences thanks to its set of mutations over the samples space specifically designed for data compression algorithms.
Data compression algorithms leverage input patterns and entropy to shrink the input into a compressed form. 
For performance reasons, their ability to search for patterns in the input is limited by different internal parameters, like lookback windows, look-ahead buffers, and history table sizes~\cite{PostgreSQL2021TOASTCompression, Euccas2021UnderstandingZlib}.
We designed the mutations that affect the sample generation process to focus on input characteristics that directly impact compression algorithm strategies and limitations towards corner cases. 

\Fuzzer mutations randomize the entropy and size of the samples that are generated.
This has an effect on the overall compressibility of sequences and literals in the sample~\cite{Euccas2021UnderstandingZlib}.
Moreover, the mutator varies the number of repeated guesses and their position in the resulting sample stressing the capability of the compression algorithm to find redundant sequences over different parts of the input.
This affects the sequence compression and can lead to corner cases, \ie where subsequent blocks to be compressed are directly marked as incompressible, as we will later show for zlib in \cref{sec:zlib_fuzzing_results}.

All of these factors together contribute to \Fuzzer's ability to amplify timing differences.

\paragrabf{Input Evolution.}
\Fuzzer follows an evolutionary approach to generate better inputs that maximize timing differences.
It generates and mutates candidate layout configurations for the attack. 
Each configuration is forwarded to the helper program that builds the requested layout, inserts the candidate guess, compresses the memory, and returns the decompression time. 
\Cref{alg:pseudocode} shows the pseudocode of the evolutionary algorithm used by \Fuzzer.

\begin{algorithm}[t]
  \footnotesize
  \SetAlgoLined
  \SetKwFunction{FEvalSample}{EvalDictAttack}
  \SetKwFunction{FSelectSamples}{SelectBest}
  \SetKwFunction{FMutate}{MutateSamples}
  \SetKwFunction{FGenerate}{GenerateRandom}
   population $\gets$ \FGenerate{}\;
   \ForEach{\upshape generation} {
    \ForEach{\upshape s $\in$ population} {
      scores[s] $\gets$ \FEvalSample{\upshape s}\;
    }
    best\_samples $\gets$ \FSelectSamples{\upshape population, scores}\;
    new\_samples $\gets$ \FMutate{\upshape best\_samples}\;
    random\_samples $\gets$ \FGenerate{}\;
    population $\gets$ best\_samples $\cup$ new\_samples $\cup$ random\_samples\;
   }
   \Return \FSelectSamples{\upshape population}\;
   \caption{\Fuzzer evolutionary pseudo-code}\label{alg:pseudocode}
\end{algorithm}

\Fuzzer iterates through different generations, with each sample having a probability of survival to the new generation that depends on its fitness score, where the fitness score is the time difference between the correct guess and the nearest incorrect one. \Fuzzer discards all the samples where the correct guess is not the fastest or slowest.
A retention factor decides the percentage of samples selected to survive among the best ones in the old generation (\SI{5}{\percent} by default).

The population for each new generation is initialized with the samples that survived the selection process and enhanced by random mutations of such samples, tuning evolution towards local optimal solutions. 
By default, \SI{70}{\percent} of the new population is generated by mutating the best samples from the previous generation.
However, to avoid trapping the fuzzer in locally optimal solutions, a percentage of completely random new samples is injected in each new generation.
\Fuzzer runs until the maximum number of generations is evaluated, and the best candidate layouts found are returned.

\subsection{Results: Fuzzing Compression Algorithms} 
\label{sec:zlib_fuzzing_results}
\paragrabf{Evaluation.}
Our test system has an Intel i7-6700K (Ubuntu 20.04, kernel 5.4.0) with a fixed frequency of \SI{4}{\giga\hertz}.
We run \Fuzzer on four compression algorithms: zlib (1.2.11), Facebook's Zstd (1.5.0), LZ4 (v1.9.3), and PGLZ in PostgreSQL (v12.7).
\Fuzzer can support new algorithms by just adding compression and decompression functions.

We run \Fuzzer with \SIx{50} epochs and a population of \SIx{1000} samples per epoch.
We set the retention factor to \SI{5}{\percent}, selecting the best 50 samples in each generation of \SIx{1000} samples.
We randomly mutate the selected samples to generate \SI{70}{\percent} of the children and add \SI{25}{\percent} of completely randomly generated layouts in the new generation.
The runtime of \Fuzzer to finish all 50 epochs was for zlib, \SI{2.46}{\hour}, for zstd, \SI{1.73}{\hour}, for LZ4, \SI{1.64}{\hour}, and for PGLZ, \SI{2.09}{\hour}.

\begin{table}[t]
\setlength{\aboverulesep}{0pt}
\setlength{\belowrulesep}{0pt}
\caption{Timing differences between correct and incorrect guesses found by \Fuzzer and the corresponding runtime.}
\label{tab:fuzzer_results}
\begin{center}
\vspace{-0.2cm}
\resizebox{\hsize}{!}{
    \begin{tabular}{|r|r|r|}
        \hline
        Algorithm                    & Max difference for correct guess (ns) & Runtime (h) \\
        \hline 
        PGLZ & 109233.25 & 2.09\\
        zlib & 71514.75 & 2.46\\
        zstd & 4239.25  & 1.73\\
        LZ4  & 2530.50  & 1.64\\
        \hline
    \end{tabular}
}
    \end{center}
\end{table}
  
\Cref{tab:fuzzer_results} lists the maximum timing differences found for \Attack on the four compression algorithms, all of which use sequence compression like LZ77. Particularly, for zlib and PGLZ, the fuzzer discovers cases with timing differences of the scale of microseconds between correct and incorrect guesses, that is, orders of magnitude higher than the manually discovered differences in \cref{sec:analysis}.
Since zlib is one of the more popular algorithms, we analyze its high latency difference in more detail below.

\paragrabf{Zlib.}
Using \Fuzzer, we found a corner case in zlib where all incorrect guesses lead to a slow code path, and the correct guess leads to a significantly faster execution time.
We ran \Fuzzer with a known prefix \ie cookie.
We observe a high timing difference of \SI{71514.75}{\nano\second} between correct and incorrect guesses, which is \textbf{3} orders of magnitude higher compared to the latency difference found manually in \cref{sec:dictionary-attack}.
This memory layout also leads to similarly high timing differences across all compression levels of zlib.
To rule out micro-architectural effects, we re-run the experiment on different systems equipped with an Intel i5-8250U, AMD Ryzen Threadripper 1920X, and Intel Xeon Silver 4208, and observe similar timing differences.

On further analysis, we observe that the corner case identified by the fuzzer is due to incompressible data.
The initial data in the page, from a uniform distribution, is primarily incompressible.
For such incompressible blocks, \Deflate algorithm can store them as raw data blocks, also called \textit{stored blocks}~\cite{Deutsch1996Deflate}.
Such blocks have fast decompression times as only a single \texttt{memcpy} operation is needed on decompression instead of the actual DEFLATE process.
In this particular corner case, the fuzzer discovered the correct guess results in such an incompressible \textit{stored block} which is faster, while an incorrect guess results in a partly compressible input which is slower, as we describe below.

\paragrabf{Correct Guess.}
On a compression for the case where the guess matches the secret, the entire guess string, \ie \texttt{cookie=SECRET}, is compressed with the secret string. All subsequent data in the input is incompressible and treated as a stored block and decompressed with a single \texttt{memcpy} operation, which is fast without any Huffman or LZ77 decoding.

\paragrabf{Incorrect Guess.}
On a compression for the case where the guess does not match the secret, only the prefix of the guess, \ie \texttt{cookie=}, is compressed with the prefix of the secret, while another longer sequence \ie \texttt{cookie=FOOBAR} leads to forming a new block. 
On a decompression, this block must now undergo the Huffman decoding  (and LZ77), which results in several table lookups, memory accesses, and higher latency.

Thus, the timing differences for the correct and incorrect guesses are amplified by the layout that \Fuzzer discovered. 
We provide more details about this layout in \cref{fig:zlib_correct_vs_incorrect} in the \cref{sec:appendix-zlib-trace} and also provide listings of the debug trace from zlib for the decompression with the correct and incorrect guesses, to illustrate the root-cause of the amplified timing differences with this layout.

\begin{tcolorbox}[boxsep=1pt,left=2pt,right=2pt,top=1pt,bottom=1pt]
  \textbf{Takeaway:} We showed that it is possible to amplify timing differences for decompression timing attacks.
  With \Fuzzer, we presented an approach to automatically find high timing differences in compression algorithms.
  The high latencies can be used for stable remote attacks.
\end{tcolorbox}

\section{Case Studies}\label{sec:casestudies}
In this section, we present three case studies showing the security impacts of the timing side channel. 
In \Cref{sec:covert_channel}, we present a local covert channel that allows hidden communication via the latency decompression.
Furthermore, we present a remote covert channel that exploits the decompression of memory objects in a PHP application using Memcached. 
We demonstrate \Attack on a PHP application that compresses secret data together with attacker-data to leak the secret bytewise.
In \Cref{sec:database}, we leak inaccessible values from a database, exploiting the internal compression of PostgreSQL. 
In \Cref{sec:vmcompression}, we show that OS-based memory compression also has timing side channels that can leak secrets. 

\subsection{Covert channel}\label{sec:covert_channel}
To evaluate the transmission capacity of memory-compression attacks, we evaluate the transmission rate for a covert channel, where the attacker controls both the sending and receiving end. 
First, we evaluate a cross-core covert channel that relies on shared memory.
This is in line with state-of-the-art work to measure the maximum capacity of the memory compression side channel~\cite{Gruss2015Template,Lipp2016,Maurice2017Hello,Yarom2014Flush,Gruss2016Flush,Guelmezoglu2015}.
The maximum capacity poses a leakage rate limit for other attacks that use memory compression as a side channel.
Our local covert channel achieves a capacity of \SI{448}{\bit/\second} ($n=100$, $\sigma=0.001\%$).

\begin{figure}[t]
    \centering
    \resizebox{\hsize}{!}{
        \resizebox{\hsize}{!}{
\begin{tikzpicture}[yscale=0.8]

\draw[fill=blue!20] (1.2,2.25) rectangle +(2.75,1.5);

\draw[fill=green!30,pattern=north west lines] (1.2,2) rectangle +(2.75,0.25) node[pos=.5,yshift=-.25cm] {\parbox{2cm}{\centering }};

\draw[pattern=north west lines] (6.5,2) rectangle +(1.,1.5) node[pos=.5] {\parbox{3.cm}{}};

\draw[fill=blue!20] (-2.5,2) rectangle +(1,1.5) node[pos=.5,yshift=-0.0cm] {\parbox{2.25cm}{}};

\draw[->,>=stealth,thick] (-1.5,2.75) to  node[midway,above]{\small Store/Update}(1.2,2.75);

\draw[->,>=stealth,thick] (6.5,2.25) to node[midway,above] {\small Fetch data}(4,2.25);

\draw [decorate,decoration={brace,amplitude=10pt},yshift=2pt]
(1,3.75) -- (4,3.75) node [midway,above,yshift=0.25cm,xshift=0.6cm]{\small\parbox{4.5cm}{Compress/Decompress}};

\node at (2.5,5) {Key-Value Store};
\node at (-2.,4.) {Sender};
\node at (7.,4.) {Receiver};

\end{tikzpicture}
}
    }
    \caption{Covert communication via a key-value store using memory compression.}
    \label{fig:covert_channel}
\end{figure}
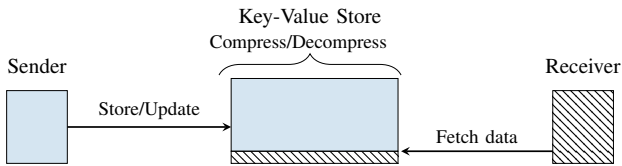

\paragrabf{Setup.}
\cref{fig:covert_channel} illustrates the covert communication between a sender and receiver via shared memory.
We create a simple key-value store that communicates via UNIX sockets.
The store takes input from a client and stores it on a \SI{4}{\kilo\byte}-aligned page.
The sender inserts a key and value into the first page to communicate with the server.
The receiver inserts a small key and value as well, which should be placed on the same \SI{4}{\kilo\byte}.
If the \SI{4}{\kilo\byte}-page is fully written, the key-value store compresses the whole page.
Compressing full \SI{4}{\kilo\byte}-page separately also occurs on filesystems like BTRFS~\cite{Btrfs2021Compression}.

\begin{figure}[t]
  \centering
  \resizebox{\hsize}{!}{
    \begin{tikzpicture}
            \begin{axis}[
            width={\hsize},
            height=3.25cm,
            xlabel={Response time [ns]},
            ylabel={Amount},
            xmax=12000,
            ymax=640,
            ymin=0,
            legend pos=outer north east,
            legend style={at={(0.65,1.0)}, anchor=north east, legend columns=1, font=\tiny},
            ]
            \addplot[thick,color=blue,densely dotted] table[x index = {0}, y index = {1}, col sep=comma]{data/results_lab05.csv};
            \addplot[thick,color=red]  table[x index = {0}, y index = {2}, col sep=comma]{data/results_lab05.csv};
            \legend{Lower entropy(0x42*4096), Higher entropy}
            \end{axis}
            
\end{tikzpicture}
  }
  \caption{Timing when decompressing a single zlib-compressed \SI{4}{\kilo\byte} page with a high-entropy in comparison to a page with a low-entropy.}
  \label{fig:histogram_zlib}
\end{figure}
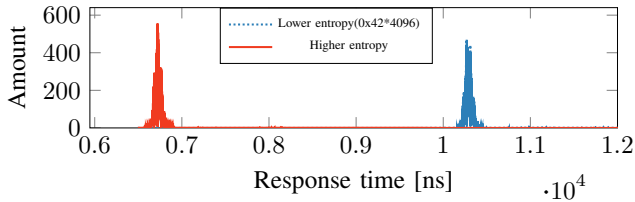

Sender and receiver agree on a time frame to send and read content.
The basic idea is to communicate via the observation on zlib that memory with low entropy, \eg 4096 times the same value, requires more time when decompressing compared to pages with a higher entropy, \eg repeating sequence number from \SIx{0} to \SIx{255}.
\cref{fig:histogram_zlib} shows the histogram of latency when decompression is triggered for both cases for the key-value on an Intel i7-6700K running at \SI{4}{\giga\hertz}.

On average, we observe a timing difference of \SI{3566.22}{\nano\second} (\SIx{14264.88} cycles, $n=100000$).
These two timing differences are easy to distinguish and enable covert communication.

\paragrabf{Transmission.}
We evaluate our cross-core covert channel by generating and sending random content from \texttt{/dev/urandom} through the memory compression timing side channel.
The sender transmits a '1'-bit by performing a store with the page that has a higher entropy \SI{4095}{\byte} in the key-value store.
Conversely, to transmit a '0'-bit, the sender compresses a low-entropy \SI{4}{\kilo\byte} page.
To trigger the compression, the receiver also stores a key-value pair in the page, wheres as \SI{1}{\byte} fills the remaining page.
The key-value store will now compress the full \SI{4}{\kilo\byte}-page, as it is fully used.
The receiver performs a fetch request from the key-value store, which triggers a decompression of the \SI{4}{\kilo\byte}-page.
To distinguish bits, the receiver measures the mean RTT of the fetch request.

\paragrabf{Evaluation.}
Our test machine is equipped with an Intel Core i7-6700K (Ubuntu 20.04, kernel 5.4.0), and all cores are running on a fixed CPU frequency of \SI{4}{\giga\hertz}.
We repeat the transmission \SIx{50} times and send per run \SI{640}{\byte}.
To reduce the error rate, the receiver fetches the receiver-controlled data \SIx{50} times and compares the average  response time against the threshold.
Our cross-core covert channel achieves an average transmission rate of \SI{448}{\bit/\second} ($n=100$, $\sigma=0.007\%$) with an error rate of \SI{0.082}{\percent} ($n=100$, $\sigma=2.85\%$).
This capacity of the unoptimized covert channel is comparable to other state-of-the-art microarchitectural cross-core covert channels that do not rely on shared memory~\cite{Wu2012,Liu2015Last,Evtyushkin2016RNG,Pessl2016,Maurice2017Hello,Semal2020One,Ragab2021,Weber2021osiris}.

\subsection{Remote Covert Channel}\label{sec:remote_covert}
We extend the scope of our covert channel to a remote covert channel.
In the remote scenario, we rely on Memcached on a web server for memory compression and decompression. 

\paragrabf{Memcached.}
Memcached is a widely used caching system for web sites~\cite{memcached_website}.
Memcached is a simple key-value store that internally uses a slab allocator. 
A slab is a fixed unit of contiguous physical memory, which is typically assigned to a certain slab class which is typically a \SI{1}{\mega\byte} region~\cite{MemcachedFreeList}.
Certain programming languages such as PHP or Python offer the option to compress memory before it is stored in Memcached~\cite{PHP2021Memcached,Python2021BMemcached}.
For PHP, memory compression is enabled per default if Memcached is used~\cite{PHP2021Memcached}.
PHP-Memcached has a threshold that decides at which size data is compressed, with the default value at \SI{2000}{\byte}.
Furthermore, PHP-Memcached computes the compression ratio to a compression factor and decides whether it stores the data compressed or uncompressed in Memcached.
The default value for the compression factor is \SIx{1.3}, \ie \SI{23}{\percent} of the space needs to be saved from the original data size to store it compressed~\cite{PHP2021Memcached}.

\paragrabf{Bypassing the Compression Factor.}
While the compression factor already introduces a timing side channel, we focus on scenarios where data is always compressed.
This is later useful for leakage on co-located data.
Intuitively, it should suffice to prepend highly-compressible data to enforce compression. 
However, we found that only prepending and adopting the offsets for secret repetitions, as for zlib, also influenced the corner case we found and the large timing difference.
We integrate prepending of compressible pages to \Fuzzer and also add the compression factor constraint to automatically discover inputs that fulfills the constraint and leads to large latencies between a correct and incorrect guess.

\paragrabf{Transmission.}
We use the page found by \Fuzzer which triggers a significantly lower decompression time to encode a `1'-bit.
Conversely, for a `0'-bit, we choose content that triggers a significantly higher decompression time.
The sender places a key-value pair for each bit index  at once into PHP-Memcached.
The receiver sends GET requests to the resource, causing decompression of the data that contains the sender content. 
The timing difference of the decompression is reflected in the RTT of the HTTP request. 
Hence, we measure the timing difference between the HTTP request being sent and the first response packet received.

\paragrabf{Evaluation.}
Our sender and receiver are equipped with an Intel i7-6700K (Ubuntu 20.04, kernel 5.4.0) and connected to the internet with a 10 Gb/s connection.
For the web server, we use a dedicated server in the Equinix~\cite{Equinix} cloud, \SIx{14} hops away from our network (over \SI{1000}{\kilo\meter} physical distance) with a \SI{10}{\giga\bit/s} connection.
The victim server is equipped with an Intel Xeon E3-1240 v5 (Ubuntu 20.04, kernel 5.4.0).
Our server runs a PHP (version 7.4) website, which allows storing and retrieving data, backed by Memcached.
We installed Memcached 1.5.22, the default version on Ubuntu 20.04. 
The PHP website is hosted on Nginx 1.18.0 with PHP-FPM.

\begin{figure}[t]
  \centering
  \resizebox{\hsize}{!}{
      \begin{tikzpicture}
            \begin{axis}[
            width={\hsize},
            height=3.25cm,
            xlabel={Response time [ns]},
            ylabel={Amount},
            xmin=400000,
            xmax=2000000,
            legend pos=outer north east,
            legend style={at={(1.0,1.0)}, anchor=north east, legend columns=1, font=\tiny},
            ]
            \addplot[thick,color=blue,densely dotted] table[x index = {1}, y index = {0}, col sep=comma]{data/remote_covert_0.csv};
            \addplot[thick,color=red]  table[x index = {1}, y index = {0}, col sep=comma]{data/remote_covert_1.csv};
            \legend{0,1}
            \end{axis}
            
\end{tikzpicture}}
  \caption{Distribution of HTTP response times for zlib-decompressed pages stored in Memcached on memory compression encoding a '1' and a '0'-bit.}
  \label{fig:remote_covert}
\end{figure}
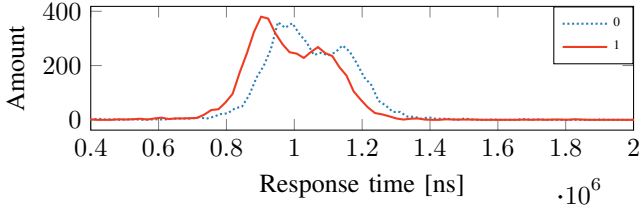

We perform a simple test where we perform \SIx{5000} HTTP requests to a PHP site that stores zlib-compressed memory in Memcached.
\Cref{fig:remote_covert} illustrates the timing difference between a `0'-bit and a `1'-bit.
The timing difference between the mean values for a `0'- and `1'-bit is \SI{61622.042}{\nano\second}.

\Cref{fig:remote_covert} shows that the two distributions overlap to a certain extent.
We call this overlap the \texttt{critical section}.
We implement an early-stopping mechanism for bits in the covert channel that clearly belong into one of the two regions below or above the critical section.
After \SIx{25} requests, we start to classify the RTT per bit.

At most, we perform \SIx{200} HTTP requests per bit.
If there are still unclassified bits, we compare the amount of requests classified as '0'-bit and those classified as '1'-bit and decide on the side with more elements.
We transmit a series of different messages of \SI{8}{\byte} over the internet.
Our simple remote covert channel achieves an average transmission rate of \SI{643.25}{\bit/\hour} ($n=20$, $\sigma=6.66\%$) at an average error rate of \SI{0.93}{\percent}.
Our covert channel outperforms the one by Schwarz~\etal\cite{Schwarz2019netspectre} and Gruss~\cite{Gruss2019page},although our attack works with HTTP instead of the more lightweight UDP sockets.
Other remote timing attacks usually do not evaluate their capacity with a remote covert channel~\cite{Zhao2009cache,Jayasinghe2010remote,Aly2013attacking,Saraswat2014,Aciicmez2007d,Aly2013attacking,VanGoethem2020Timeless}.

\subsection{Remote Attack on PHP-Memcached}\label{sec:attack_eval}
Using our building blocks to perform \Attack and the remote covert channel, we perform a remote attack on PHP-Memcached to leak secret data from a server over the internet.
We assume a memory layout where secret memory is co-located to attacker-controlled memory, and the overall memory region is compressed.

\paragrabf{Attack Setup.}
We use the same PHP-API as in~\Cref{sec:remote_covert} that allows storing and retrieving data.
We run the attack using the same server setup as used for the remote covert channel attacking a dedicated server in the Equinix~\cite{Equinix} cloud \SIx{14} hops away from our network.
We define a \SI{6}{\byte} long secret (\texttt{SECRET}) with a \SI{7}{\byte} long prefix (\texttt{cookie=}) and prepend it to the stored data of users.
PHP-Memcached will compress the data before storing it in Memcached and decompress it when accessing it again.
For each guess, the PHP application stores the uploaded data to a certain location in Memcached.
On each data fetch, the PHP application decompresses the secret data together with the co-located attacker-controlled data and then responds only-the attacker-controlled data.
The attacker measures the RTT per guess and distinguishes the timing differences between the guesses.
In the case of zlib, the guess with the fastest response time is the correct guess.
We demonstrate a byte-wise leakage of an arbitrary secret and also a dictionary attack with a dictionary size of 100 guesses.

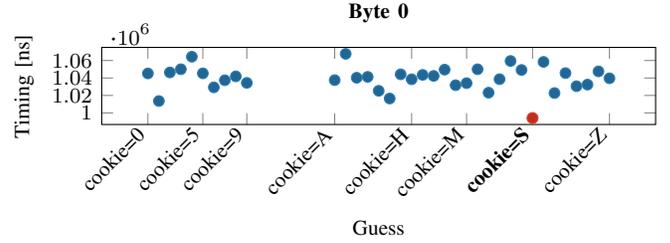
\begin{figure}[t]
  \centering
  \resizebox{\hsize}{!}{
      \begin{tikzpicture}
\begin{axis}[
title={\textbf{Byte 0}},
style={font=\footnotesize},
xlabel={Guess},
ylabel={Timing [ns]},
y label style={align=center,text width=2cm},
width=\hsize,
height=2.6cm,
xtick={48,53,57,65,72,77,83,90},
xticklabels={cookie=0,cookie=5,cookie=9,cookie=A,cookie=H,cookie=M,\textbf{cookie=S},cookie=Z},
x tick label style={rotate=45,anchor=east}
]

\addplot+[mark=*,draw=none] table[x=Secret,y=Prev_TS_Frame,col sep=comma] {data/memcached_bytewise.csv};
\addplot+[mark=*,only marks,error bars/.cd,y dir=both,y explicit]
        table[x=Secret,y=Prev_TS_Frame,col sep=comma,
        restrict expr to domain={\thisrow{Secret}}{83:83}
        ]{data/memcached_bytewise.csv}; 
\end{axis}
\end{tikzpicture}}
  \caption{Distribution of the response time for the correct guess (S) and the incorrect guesses (0-9, A-R, T-Z) for the first byte leaked in the byte-wise leakage of the secret from PHP-Memcached. Similar trends are observed for subsequent bytes leaked as shown in \cref{sec:appendix-bytewise-leakage}.}
  \label{fig:memcached_attack_bytewise}
\end{figure}

\paragrabf{Evaluation.}
For the byte-wise attack, we assume each byte of the secret is from ``A-Z,0-9" (36 different options).
For each of the 6 bytes to be leaked, we generate a separate memory layout using \Fuzzer that maximizes the latency between the guesses.
We repeat the experiment \SIx{20} times.
On average, our attack leaks the entire \SI{6}{\byte} secret string in \SI{31.95}{\minute} ($n=20$, $\sigma=8.48\%$).
Since the latencies between a correct and incorrect guess are in the microseconds range, we do not observe false positives with our approach.
\Cref{fig:memcached_attack_bytewise} shows the median response time for each guess in the first iteration (leaking the first byte) as a representative example.
It can be seen that the response time for the correct guess (\texttt{S}) for the first byte is significantly faster than the incorrect guesses. We observe similar results for the remaining bytes (\texttt{E,C,R,E,T}). 

\paragrabf{Dictionary attack.}
While bytewise leakage is the more generic and practical approach for decompression timing attacks, there might be use cases where also a dictionary attack is applicable.
For instance, if the attacker would try the top $n$ most common usernames or passwords, which are publicly available.
Another use case for dictionary attacks could be to profile if certain images are forbidden in a database used for digital rights management.

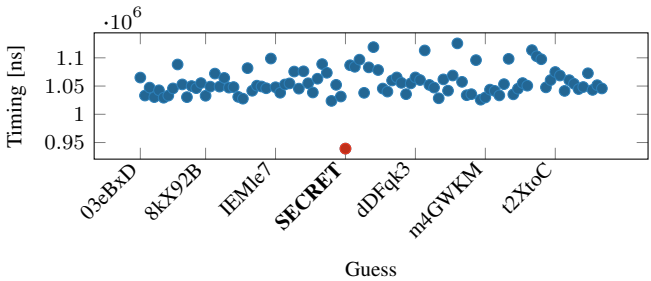
\begin{figure}[t]
  \centering
  \resizebox{\hsize}{!}{
      \begin{tikzpicture}
\begin{axis}[
style={font=\footnotesize},
xlabel={Guess},
ylabel={Timing [ns]},
y label style={align=center,text width=2cm},
width=\hsize,
height=3.25cm,
xtick={1,15,30,45,60,75,90},
xticklabels={03eBxD,8kX92B,IEMle7,\textbf{SECRET},dDFqk3,m4GWKM,t2XtoC},
x tick label style={rotate=45,anchor=east}
]

\addplot+[mark=*,draw=none] table[x=Index,y=Time,col sep=semicolon] {data/memcached_attack_median.csv};
\addplot+[mark=*,only marks,error bars/.cd,y dir=both,y explicit]
        table[x=Index,y=Time,col sep=semicolon,
        restrict expr to domain={\thisrow{Index}}{45:45}
        ]{data/memcached_attack_median.csv};
\end{axis}
\end{tikzpicture}}
  \caption{Distribution of the response time for the correct guess (SECRET) and the 99 other incorrect guesses in dictionary attack on PHP-Memcached. Note that we only label SECRET and 6 out of 99 incorrect guesses on the X-axis for brevity.}
  \label{fig:memcached_attack}
\end{figure}

We perform a dictionary attack over the network using randomly generated \SIx{100} guesses, including the correct guess.
We repeat the experiment \SIx{20} times.
For each run, we re-generate the guesses and randomly shuffle the order of the guesses per request iteration to not create a potential bias.
On average, our attack leaks the secret string in \SI{14.99}{\minute} ($n=20$, $\sigma=16.27\%$).
To classify the correct guess, we use the same early stopping technique as was used for the covert channel.
Since the timing difference is in the microseconds range, we do not observe false positives with our approach.
\Cref{fig:memcached_attack} illustrates the median of the response times for each guess from one iteration.
\Cref{fig:memcached_all_bytes} (in \cref{sec:appendix-bytewise-leakage}) shows the latencies for all leaked bytes.
It can be clearly seen that the response time for the correct guess (\texttt{SECRET}) is significantly faster.

\begin{tcolorbox}[boxsep=1pt,left=2pt,right=2pt,top=1pt,bottom=1pt]
  \textbf{Takeaway:} We show that a PHP application using Memcached to cache blobs hosted on Nginx enables covert communication with a transmission rate of \SI{643.25}{\bit/\hour}.
  If secret data is stored together with attacker-controlled data, we demonstrate a remote memory-compression attack on Memcached leaking a \SI{6}{\byte}-secret bytewise in \SI{31.95}{\minute}.
  We further demonstrated that the attack can also be used for dictionary attacks.
\end{tcolorbox}

\subsection{Leaking Data from Compressed Database}\label{sec:database}
In this section, we show that an attacker can also exploit compression in databases to leak inaccessible information, \ie from the internal database compression of PostgreSQL.

\paragrabf{PostgreSQL Data Compression.} 
PostgreSQL is a wide\-spread open-source relational database system using the SQL standard. 
PostgreSQL maintains tuples saved on disk using a fixed page size of commonly \SI{8}{\kilo\byte}, storing larger fields compressed and possibly split into multiple pages.

By default, variable-length fields that may produce large values, \eg \texttt{TEXT} fields, are stored compressed.
PostgreSQL's transparent compression is known as \texttt{TOAST} (The Oversized-Attribute Storage Technique) and uses a fast LZ-family compression, \texttt{PGLZ}~\cite{PostgreSQL2021TOASTCompression}.
By default, data in a cell is actually stored compressed if such a form saves at least \SI{25}{\percent} of the uncompressed size to avoid wasting decompression time if it is not worth it.
Indeed, data stored uncompressed is accessed faster than data stored compressed since the decompression algorithm is not executed. 
Moreover, the decompression time depends on the compressibility of the underlying data.
A potential attack scenario for structured text in a cell is where JSON documents are stored and compressed within a single cell, and the attacker controls a specific field within the document.
While our focus is restricted to cell-level compression, note that compressed columnar storage~\cite{Citus2021Columnar,Microsoft2021Concepts,Buckenhofer2021PostgreSQL}, a non-default extension available in PostgreSQL, may also be vulnerable to decompression timing attacks.

\paragrabf{Attack Setup.}
To assess the feasibility of an attack, we use a local database server and access two differently compressed rows with a Python wrapper using the \texttt{psycopg2} library. 
The first row contains \SIx{8192} characters of highly compressible data, while the second one \SIx{8192} characters of random incompressible data. 
Both rows are stored in a table as \texttt{TEXT} data and accessed 1000 times. 
The median for the number of clock cycles required to access the compressible row is \SIx{249031}, while for the uncompressed one is \SIx{221000}, which makes the two accesses distinguishable. 
On our \SI{4}{\giga\hertz} CPUs, this is a timing difference of \SI{7007.75}{\nano\second}.
We use \Fuzzer to amplify these timing differences and demonstrate a bytewise leakage of a secret and also a dictionary attack.

\paragrabf{Leaking First Byte.}
For the bytewise leeakage of the secret, we first create a memory layout to leak the first byte using \Fuzzer against a standalone version of PostgreSQL's compression library, using a similar setup as the previous Memcached attack. 
A key difference in the use of \Fuzzer with PostgreSQL is that the helper program measuring the decompression time returns a time of 0 when the input is not compressed, \ie the data compressed with PGLZ does not save at least \SI{25}{\percent} of the original size.
\Fuzzer found a layout that sits exactly at the corner case where a correct guess in the secret results in a compressed size that saves \SI{25}{\percent} of the original size. 
Hence, a correct guess is saved compressed, while for any wrong guess, the threshold is not met, and the data is saved uncompressed. 
This is because the correct guess characters match the secret bytes and result in a higher compression rate. 

\paragrabf{Leaking Subsequent Bytes with Secret Shifting.}
We observed that one good layout is enough and can be reused for bytewise leakage in PGLZ.
The prefix can be shifted by one character to the left by a single character, \ie from ``cookie=S" to ``ookie=SE", to accommodate an additional byte for the guess.
This shifting allows bytewise leakage with the same memory layout.
Note, that the same approach did not work on \Deflate.
We leave it as future work, to mount such a shifting approach to other compression algorithms.

\paragrabf{Evaluation.}
We perform a remote decompression timing attack against a Flask~\cite{Flask2021Homepage} web server that uses a PostgreSQL database to store user-provided data. 
We used the same Equinix cloud-server setup as used for the Memcached remote attack (\cf \Cref{sec:attack_eval}). 
The server runs Python 3.8.5 with Flask version 2.0.1 and PostgreSQL version 12.7. 

We define a \SI{6}{\byte} long-secret with a \SI{7}{\byte} prefix that the server colocates with attacker-provided data in a cell of the database through POST requests. 
The secret is never shown to the user. 
Using the layout found by \Fuzzer, the entry in the database is stored compressed only when the secret matches in the provided data.
A second endpoint in the server accesses the database to read the data without returning the secret to the attacker.

\begin{figure}[t]
  \centering
  \resizebox{\hsize}{!}{
      \begin{tikzpicture}
\begin{axis}[
title={\textbf{Byte 0}},
style={font=\footnotesize},
xlabel={Guess},
ylabel={Timing [ns]},
y label style={align=center,text width=2cm},
width=\hsize,
height=2.8cm,
xtick={48,53,57,65,72,77,83,90},
xticklabels={cookie=0,cookie=5,cookie=9,cookie=A,cookie=H,cookie=M,\textbf{cookie=S},cookie=Z},
x tick label style={rotate=45,anchor=east}
]

\addplot+[mark=*,draw=none] table[x=Secret,y=Prev_TS_Frame,col sep=comma] {data/postgres_bytewise_0.csv};
\addplot+[mark=*,only marks,error bars/.cd,y dir=both,y explicit]
        table[x=Secret,y=Prev_TS_Frame,col sep=comma,
        restrict expr to domain={\thisrow{Secret}}{83:83}
        ]{data/postgres_bytewise_0.csv}; 
\end{axis}
\end{tikzpicture}}
  \caption{Distribution of the response time for the first-byte guesses in the remote PostgreSQL attack leaking a secret bytewise, with the correct guess (S) and incorrect guesses (0-9, A-R, T-Z). Similar trends are observed for subsequent bytes leaked as shown in \cref{sec:appendix-bytewise-leakage}.}
  \label{fig:postgres_attack_bytewise}
\end{figure}
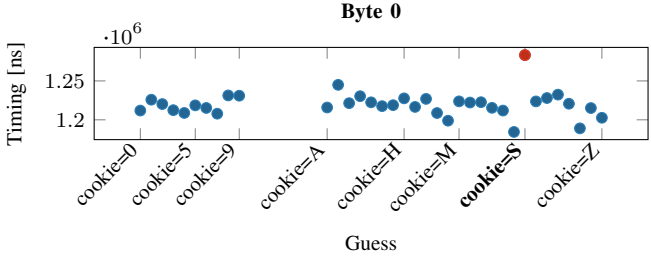

The attack leaks bytewise over the internet by guessing again in the character set ``A-Z,0-9'' (\SIx{36} possibilities per character), including the correct one.
We repeat the attack \SIx{20} times.
The average time required to determine the guess with the highest latency (that the server had to decompress before returning) is \SI{17.84}{\minute} (\SI{2.97}{\minute/\byte}) ($n=20$, $\sigma=2.3\%$).
\Cref{fig:postgres_attack_bytewise} illustrates the median of the response times for a single guess in an example iteration, showing how the correct guess results in a slower response.

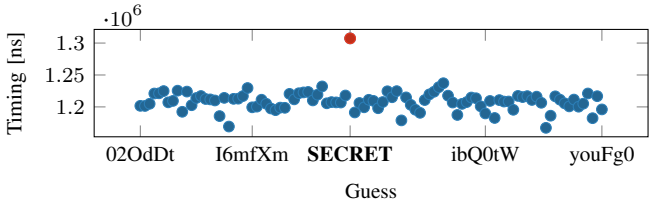
\begin{figure}[t]
  \centering
  \resizebox{\hsize}{!}{
      \begin{tikzpicture}
\begin{axis}[
style={font=\footnotesize},
xlabel={Guess},
ylabel={Timing [ns]},
y label style={align=center,text width=2cm},
width=\hsize,
height=3cm,
xtick={1,25,46,75,100},
xticklabels={02OdDt,I6mfXm,\textbf{SECRET},ibQ0tW,youFg0}
]

\addplot+[mark=*,draw=none] table[x=Index,y=Time,col sep=semicolon] {data/postgres_attack_median.csv};
\addplot+[mark=*,only marks,error bars/.cd,y dir=both,y explicit]
        table[x=Index,y=Time,col sep=semicolon,
        restrict expr to domain={\thisrow{Index}}{46:46}
        ]{data/postgres_attack_median.csv};
\end{axis}
\end{tikzpicture}}
  \caption{Distribution of the response time in the remote PostgreSQL dictionary attack, for the correct guess
(SECRET) and the 99 other incorrect guesses (we only label 4 out of 99 incorrect guesses and SECRET on the X-axis for brevity).
}
  \label{fig:postgres_attack}
\end{figure}

\paragrabf{Dictionary attack.}
We also performed a dictionary attack over the internet with \SIx{100} randomly generated guesses, including the correct one, and repeated the experiments \SIx{20} times.
We measured the time required for the server to reply to a request in the second endpoint by observing the network packets arriving at our machine with the same setup as the remote Memcached attack, repeating each request \SIx{100} times.
The average time required to brute-force all the guesses and select the guess with the highest latency (that the server had to decompress before returning) is \SI{8.28}{\minute} ($n=20$, $\sigma=0.003\%$).
\Cref{fig:postgres_attack} illustrates the median of the response times for each guess in an example iteration, showing how the correct guess results in a slower response.
Again, the timing difference is clearly distinguishable over \SI{14} hops in the internet.

\begin{tcolorbox}[boxsep=1pt,left=2pt,right=2pt,top=1pt,bottom=1pt]
  \textbf{Takeaway:} Secrets can be leaked from databases due to timing differences caused by PostgreSQL's transparent compression, if an application stores untrusted data with secrets in the same cell of a database.
With a Flask appplication using a PostgreSQL database, we show that an attacker can infer when its inserted data matches the secret based on response times, which are influence by the decompression timing.
Our decompression timing attack on PostgreSQL leaks a \SI{6}{\byte} secret bytewise over the network in \SI{17.84}{\minute}.
\end{tcolorbox}

\subsection{Attacking OS Memory Compression}\label{sec:vmcompression}
In this section, we show how memory compression in modern OSs can introduce exploitable timing differences.
In particular, we demonstrate a dictionary-based attack to leak secrets from compressed pages in ZRAM, the Linux implementation of memory compression.

\paragrabf{Background.}
Memory compression is a technique used in many modern OSs, \eg  Linux~\cite{Larabel2020Fedorazram}, Windows~\cite{Hoffman2017Windowsmemcomp}, or MacOS~\cite{Denny2020Chromeoszram}.
Similar to traditional swapping, memory compression increases the effective memory capacity of a system.
When processes require more memory than available, the OS can transparently compress unused pages in DRAM to ensure they occupy a smaller footprint in DRAM rather than swapping them to disk.
This frees up memory while still allowing the compressed pages to be accessed from DRAM. 
Compared to disk I/O, DRAM access is an order of magnitude faster, and even with the additional decompression overhead, memory compression is significantly faster than swapping.
Hence, memory compression can improve the performance despite the additional CPU cycles required for compression and decompression.
Such memory compression is adopted in the Linux kernel (since kernel version 3.14) in a module called ZRAM~\cite{Gupta2021Zram}, which is enabled by default on Fedora~\cite{Larabel2020Fedorazram} and Chrome OS~\cite{Denny2020Chromeoszram}.

\subsubsection{Characterizing Timing Differences in ZRAM} \label{sec:zram_characterization}
To understand how memory compression can be exploited, we characterize its behavior in ZRAM.
On Linux systems, ZRAM appears as a DRAM-backed block device. 
When pages need to be swapped to free up memory, they are instead compressed and moved to ZRAM.
Subsequent accesses to data in ZRAM result in a page fault, and the page is decompressed from ZRAM and copied to a regular DRAM page for use again. 
We show that the time to access data from a ZRAM page depends on its compressibility and thus the data values.

\begin{table}[t]
	\setlength{\aboverulesep}{0pt}
	\setlength{\belowrulesep}{0pt}
	\caption{Mean latency of accesses to ZRAM. Distinguishable timing differences exist based on data compressibility in the pages ($n=500$ and 6\% of samples removed as outliers with more than an order of magnitude higher latency).}
	\label{tab:zram_algo_timing}
	\vspace{-0.1cm}	
	\begin{center}
		\resizebox{\hsize}{!}{
			\begin{tabular}{|l|r|r|r|}
				\hline				
				\multirow{2}{*}{Algorithm}   & \multirow{2}{*}{Incompressible (ns)} & Partly~\qquad\qquad  & Fully~\qquad\qquad \\
				                             &               & Compressible (ns)  & Compressible (ns) \\				
				\hline \addlinespace[0.15cm]
				deflate & \SIx{1763} ($\pm12\%$) & \SIx{12208} ($\pm2\%$) & \SIx{1551} ($\pm12\%$)  \\
				842 & \SIx{1789} ($\pm11\%$) & \SIx{8785} ($\pm2\%$) & \SIx{1556} ($\pm10\%$)       \\
				lzo & \SIx{1684} ($\pm\phantom{0}9\%$) & \SIx{4866} ($\pm4\%$) & \SIx{1479} ($\pm12\%$)        \\
				lzo-rle & \SIx{1647} ($\pm\phantom{0}9\%$) & \SIx{4751} ($\pm4\%$) & \SIx{1453} ($\pm12\%$)    \\
				zstd & \SIx{1857} ($\pm10\%$) & \SIx{2612} ($\pm9\%$) & \SIx{1674} ($\pm11\%$)      \\
				lz4 & \SIx{1710} ($\pm11\%$) & \SIx{1990} ($\pm7\%$) & \SIx{1470} ($\pm10\%$)       \\
				lz4hc & \SIx{1746} ($\pm\phantom{0}9\%$) & \SIx{2091} ($\pm9\%$) & \SIx{1504} ($\pm11\%$)      \\\addlinespace[0.1cm]
				\hline 
			\end{tabular}
		}
	\end{center}
\end{table}

We characterize the latency of accessing data from ZRAM pages with different entropy levels: pages that are \textit{incompressible} (with random bytes), \textit{partially-compressible} (random values for 2048 bytes and a fixed value repeated for the remaining 2048 bytes), and \textit{fully-compressible} (a fixed value in each of the 4096 bytes).
We ensure a page is moved to ZRAM by accessing more memory than the memory limit allows. 
To ensure fast run times for the proof of concept, we allocate the process to a \textit{cgroup} with a memory limit of a few megabytes.
We measure the latency for accessing a 8-byte word from the page in ZRAM, and repeat this process 500 times.  

\cref{tab:zram_algo_timing} shows the mean latency for ZRAM accesses for different ZRAM compression algorithms on an Intel i7-6700K (Ubuntu 20.04, kernel 5.4.0).
The latency for accesses to ZRAM is much higher for partially compressible pages (with lower entropy) compared to incompressible pages (with higher entropy) for all the compression algorithms.
This is because the process of moving compressed ZRAM pages to regular memory on an access requires additional calls to functions that decompress the page; on the other hand, ZRAM pages that are stored uncompressed do not require these function calls (\cref{sec:appendix-zram-ftrace} provides a trace of the kernel function calls on ZRAM accesses and their execution times to illustrate this difference).
The largest timing difference is observed for \texttt{deflate} algorithm (close to \SI{10000}{\nano\second}) and \texttt{842} algorithm (close to \SI{7000}{\nano\second}); moderate timing differences are observed for \texttt{lzo} and \texttt{lzo-rle} (close to \SI{1000}{\nano\second}), and \texttt{zstd} (close to \SI{750}{\nano\second}); the smallest timing difference is seen for \texttt{lz4} and \texttt{lz4hc} (close to \SI{250}{\nano\second}).
These timing differences largely correspond with the raw decompression latency for these algorithms, some of which are measured in Table~\ref{tab:evaluated_algorithms_decomp}.

Accesses to a fully-compressible page in ZRAM, \ie a page containing the same byte repeatedly, are faster (by \SI{200}{\nano\second}) than accesses to an incompressible page for all the compression algorithms. 
This is because ZRAM stores such pages with a special encoding as a single-byte (independent of the compression algorithm) that only requires reading a single byte from ZRAM on an access to such a page.

\subsubsection{Leaking Secrets via ZRAM Decompression Timings}
In this section, we exploit timing differences between accesses to a partially compressible and incompressible page in ZRAM (using \texttt{deflate} algorithm) to leak secrets.
Like previous PoCs, we use \Fuzzer to generate data layouts that maximize the decompression-timing difference for the correct guess, while leaking a secret byte-wise.

\textbf{Attack Setup.} We demonstrate byte-wise leakage attack on a program with a 4KB page stored in ZRAM containing both a secret value and attacker-controlled data, as is common in many applications like databases.
To determine optimal data layouts an attacker might use, we combine this program with \Fuzzer. 
With a known secret value, \Fuzzer runs the program with the attacker guessing each byte position successively. 
For each byte position, \Fuzzer measures the decompression time for the page from ZRAM with different attacker data layouts and different attacker guesses, and the layout maximizing the timing difference when the attacker's guessed-byte matches the secret-byte is returned. 
In such an optimal layout, when the attacker's guess matches the secret-byte, the page entropy reduces (page is partially compressible) and ZRAM decompression takes longer; and for all other guesses, the entropy is high and ZRAM decompression is fast.
We repeat this process to generate optimal data layouts for each byte position.
Note that this optimal data layout only relies on the number of repetitions of the guess and the relative position of the guessed data and the secret (a property of the compression algorithm), and is applicable with any data values. 

Using these attacker data-layouts, we perform the byte-wise leakage of an unknown secret. 
At each step, the attacker makes one-byte guesses (0-9, A-Z) and then denotes the guess with the highest latency to be the ``correct'' guess. It appends this value to its guess-string and then guesses the next byte-position. We repeat this attack for 100 random secret strings.

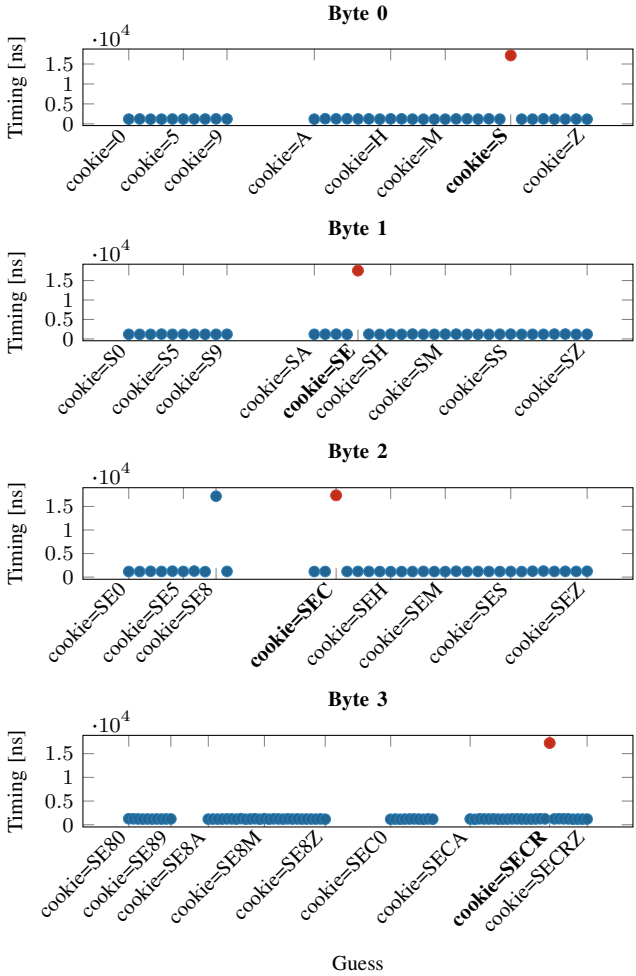
\begin{figure}[t]
  \centering
	\begin{subfigure}{\hsize}%
        \begin{tikzpicture}
\begin{axis}[
title={\textbf{Byte 0}},
style={font=\footnotesize},
ylabel={Timing [ns]},
y label style={align=center,text width=2cm},
width=\hsize,
height=2.6cm,
xtick={48,53,57,65,72,77,83,90},
xticklabels={cookie=0,cookie=5,cookie=9,cookie=A,cookie=H,cookie=M,\textbf{cookie=S},cookie=Z},
x tick label style={rotate=45,anchor=east}
]

\addplot+[mark=*,draw=none] table[x=Char,y=Latency_ns,col sep=comma] {data/zram_bytewise_0.csv};
\addplot+[mark=*,only marks,error bars/.cd,y dir=both,y explicit]
        table[x=Char,y=Latency_ns,col sep=comma,
        restrict expr to domain={\thisrow{Char}}{83:83}
        ]{data/zram_bytewise_0.csv};
\end{axis}
\end{tikzpicture}%
    \end{subfigure}
    \begin{subfigure}{\hsize}%
        \begin{tikzpicture}
\begin{axis}[
title={\textbf{Byte 1}},
style={font=\footnotesize},
ylabel={Timing [ns]},
y label style={align=center,text width=2cm},
width=\hsize,
height=2.6cm,
xtick={48,53,57,65,69,72,77,83,90},
xticklabels={cookie=S0,cookie=S5,cookie=S9,cookie=SA,\textbf{cookie=SE},cookie=SH,cookie=SM,cookie=SS,cookie=SZ},
x tick label style={rotate=45,anchor=east}
]

\addplot+[mark=*,draw=none] table[x=Char,y=Latency_ns,col sep=comma] {data/zram_bytewise_1.csv};
\addplot+[mark=*,only marks,error bars/.cd,y dir=both,y explicit]
        table[x=Char,y=Latency_ns,col sep=comma,
        restrict expr to domain={\thisrow{Char}}{69:69}
        ]{data/zram_bytewise_1.csv}; 
\end{axis}
\end{tikzpicture}%
    \end{subfigure}
    \begin{subfigure}{\hsize}%
    	\begin{tikzpicture}
\begin{axis}[
title={\textbf{Byte 2}},
style={font=\footnotesize},
ylabel={Timing [ns]},
y label style={align=center,text width=2cm},
width=\hsize,
height=2.8cm,
xtick={48,53,56,67,72,77,83,90},
xticklabels={cookie=SE0,cookie=SE5,cookie=SE8,\textbf{cookie=SEC},cookie=SEH,cookie=SEM,cookie=SES,cookie=SEZ},
x tick label style={rotate=45,anchor=east}
]

\addplot+[mark=*,draw=none] table[x=Char,y=Latency_ns,col sep=comma] {data/zram_bytewise_2.csv};
\addplot+[mark=*,only marks,error bars/.cd,y dir=both,y explicit]
        table[x=Char,y=Latency_ns,col sep=comma,
        restrict expr to domain={\thisrow{Char}}{67:67}
        ]{data/zram_bytewise_2.csv}; 
\end{axis}
\end{tikzpicture}%
    \end{subfigure}
    \begin{subfigure}{\hsize}%
        \begin{tikzpicture}
\begin{axis}[
title={\textbf{Byte 3}},
style={font=\footnotesize},
 xlabel={Guess},
ylabel={Timing [ns]},
y label style={align=center,text width=2cm},
width=\hsize,
height=2.8cm,
xtick={48,57,65,77, 90, 104, 121, 138, 146},
xticklabels={cookie=SE80,cookie=SE89,cookie=SE8A,cookie=SE8M,cookie=SE8Z,cookie=SEC0,cookie=SECA,\textbf{cookie=SECR},cookie=SECRZ},
x tick label style={rotate=45,anchor=east}
]

\addplot+[mark=*,draw=none] table[x=Char,y=Latency_ns,col sep=comma] {data/zram_bytewise_3.csv};
\addplot+[mark=*,only marks,error bars/.cd,y dir=both,y explicit]
        table[x=Char,y=Latency_ns,col sep=comma,
        restrict expr to domain={\thisrow{Char}}{138:138}
        ]{data/zram_bytewise_3.csv}; 
\end{axis}
\end{tikzpicture}%
    \end{subfigure}

  \caption{Times for guesses (0-9, A-Z) for each of the first four bytes (S, E, C, R) in a 6-byte secret (cookie=\textbf{SECRET}) leaked byte-wise from ZRAM. The highest times correspond to the secret-byte value (shown in red). The last two bytes (E, T) are also successfully leaked similarly and shown in  \cref{sec:appendix-bytewise-leakage}.}
  \label{fig:zram_attack_bytewise_Byte1234}
\end{figure}

\textbf{Evaluation.}
\Cref{fig:zram_attack_bytewise_Byte1234} shows the bytewise leakage for a secret value (cookie=\texttt{SECRET}), with the decompression times for guesses of the first four bytes depicted in each of the graphs. For each byte, among guesses of (0-9,A-Z), the  highest decompression time successfully leaks the secret byte value (shown in red). For example, for byte-0, the highest time is for cookie=\texttt{S} and similarly for byte-1, it is cookie=\texttt{SE}. For byte-2, we observe a false-positive, cookie=\texttt{SE8}, which also has a high latency along with the correct guess cookie=\texttt{SEC}. But in the subsequent byte-3, with both these strings used as prefixes for the guesses, the false-positives are eliminated and cookie=\texttt{SECR} is obtained as the correct guess. The next two bytes are also successfully leaked to fully obtain cookie=\texttt{SECRET}, as shown in the \Cref{fig:zram_attack_bytewise_byte56} in \Cref{sec:appendix-bytewise-leakage}. Our attack successfully completes in less than 5 minutes.

Over 100 randomly generated secrets, we observe that 90\% of secrets are exactly leaked successfully. In 9 out 10 remaining cases, we narrow down the secret to within four 6-byte candidates (due to false-positives) and in the last case we recover only 4 out of 6-bytes of the secret (as the false-positives grows considerably beyond this). These false-positives arise due to the data-layouts generated not being as robust as in previous PoCs. \Fuzzer with ZRAM is a few orders of magnitude slower (almost 0.03x the speed) compared to iterations with raw algorithms studied in \cref{sec:zlib_fuzzing_results}, so it is unable to explore as large a search space as previous PoCs. This is because moving a page to ZRAM and compressing it requires accessing sufficient memory to swap the page out, which is slower than just executing the compression algorithm. But such false positives can be easily addressed by using multiple data-layouts per byte (each layout causes different false-positives), or by fuzzing for a longer duration to generate more robust data-layouts. The fuzzing speed can also be increased by implementing the ZRAM compression algorithm (a fork of zlib) inside \Fuzzer, so this is not a fundamental limitation.

\begin{figure}[t]
	\centering
	\resizebox{\hsize}{!}{
		\begin{tikzpicture}
\begin{axis}[
style={font=\footnotesize},
xlabel={Guess},
ylabel={Timing [ns]},
y label style={align=center,text width=2cm},
width=\hsize,
height=3.25cm,
xtick={1,25,45,75,100},
xticklabels={9Q7Lwt,lMP6Pv,\textbf{SECRET},0NSQFp,e4gvLN},
]

\addplot+[mark=*,draw=none] table[x=index,y=latency,col sep=semicolon] {data/zram_dict_attack_medians.csv};
\addplot+[mark=*,only marks,error bars/.cd,y dir=both,y explicit]
        table[x=index,y=latency,col sep=semicolon,
        restrict expr to domain={\thisrow{index}}{45:45}
        ]{data/zram_dict_attack_medians.csv};
\end{axis}
\end{tikzpicture}}
	\caption{Median latency for ZRAM accesses in dictionary attack, with correct guess \textbf{SECRET} and 99 incorrect guesses (we only label 4 out of 99 incorrect guesses on X-axis for brevity).}
	\label{fig:zram_dict_attack_latencies}
\end{figure}
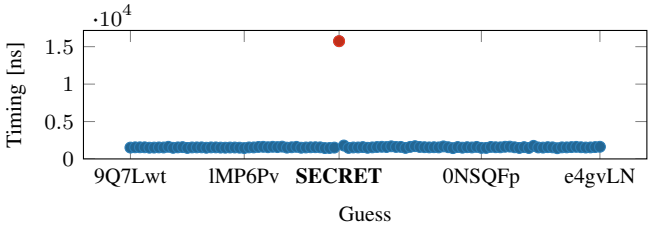

\paragrabf{Dictionary Attack.}
For completeness, we also demonstrate a dictionary attack where the secret is guessed from a range of 100 possible guesses (dictionary), similar to previous PoCs.
We assume the target page has a secret embedded the top-half and the bottom half of the page is attacker-controlled and filled with repeated guesses from a dictionary. 
\cref{fig:zram_dict_attack_latencies} shows the median latency for attacker's data access from ZRAM (over 10 observations) for different guesses from the dictionary.
The guesses consist of 99 randomly generated strings and the secret (cookie=\texttt{SECRET}). 
We observe a high timing for decompression when the guess matches the secret as the page entropy is lowered (and the page compressed in ZRAM), thus leaking the secret; for all other guesses  the decompression is fast. 
This dictionary attack completes in less than 10 minutes. But in general, the duration depends on the time required to force a target page into ZRAM, depending on the physical memory limit for a target process. 

\begin{tcolorbox}[boxsep=1pt,left=2pt,right=2pt,top=1pt,bottom=1pt]
  \textbf{Takeaway:} We show that even if an application does not explicitly use compression, its data may still get compressed by the OS due to memory compression.
Our case study of ZRAM shows that all compression algorithms in ZRAM have timing differences of \SIrange{250}{10400}{\nano\second} for accesses to compressed versus uncompressed pages.
We also show that if secrets are co-located with attacker-controlled data in the same page and compressed with memory compression, they can be leaked in minutes.
\end{tcolorbox}

\section{Mitigations}\label{sec:discussion}

\paragrabf{Disabling LZ77.}
The most rudimentary solution is to completely disable compression or at least disable the LZ77 part.
Karakostas~\etal\cite{Karakostas2016CTX} showed for web pages that this adds an overhead between \SI{91}{\percent} and \SI{500}{\percent}.
Furthermore, attacks have not been studied well enough on symbol compression to provide any security guarantees.

\paragrabf{Masking.}
Karakostas~\etal\cite{Karakostas2016CTX} presented a generic defense technique called Context Transformation Extension (CTX).
The general idea is to use context-hiding to protect secrets from being compressed with attacker-controlled data.
Data is permuted on the server-side using a mask, and on the client-side, an inverse permutation is performed (JavaScript library).
The overhead compared to the original algorithms decrease with the number of compressed data~\cite{Karakostas2016CTX}.

\paragrabf{Randomization.}
Yang~\etal\cite{Yang2019Lempel} showed an approach with randomized input to mitigate compression side-channel attacks.
The service would require adding an additional amount of random data to hide the size of the compressed memory.
However, as the authors also show, randomization-based approaches can be defeated with at the expense of a higher execution time.
Also, Karaskostas~\etal\cite{Karakostas2016CTX} showed that size randomization is ineffective against memory compression attacks.
It is also unclear if size randomization mitigates the timing-based side channel of the memory decompression. 

\paragrabf{Keyword protection.}
Zieliński demonstrated an implementation of \Deflate called SafeDeflate~\cite{Zielinski2016Safedeflate}.
SafeDeflate mitigates memory compression attacks by splitting the set of keywords into subsets of sensitive and non-sensitive keywords.
Depending on the completeness of the sensitive keyword list, this approach can be considered as secure, but there is no guaranteed security.
As Paulsen~\etal\cite{Paulsen2019Debreach} mention, it is easy to overlook a corner case.
Furthermore, this approach leads to a loss of compression ratio of about \SIrange{200}{400}{\percent}~\cite{Karakostas2016CTX}.

\paragrabf{Taint tracking.}
Taint tracking tries to trace the data flow and mark input sources and their sinks.
Paulsen~\etal\cite{Paulsen2019Debreach} use taint analysis to track the flow of secret data before feeding data into the compression algorithm.
Their tool, Debreach, is about 2-5 times faster than SafeDeflate~\cite{Paulsen2019Debreach}.
However, this approach is, so far, only compatible with PHP.
Furthermore, for such an approach, the developer needs to flag the sensitive input which is being tracked.

The probably best strategy in practice is to avoid sensitive data being compressed with potential attacker-controlled data.
The aforementioned mitigations focus on mitigating compression ratio side channels. 
As the compression and decompression timings are not constant, a timing side channel is harder to mitigate, \eg remote attacks might be mitigated by adding artificial noise or using network packet inspection to detect attacks~\cite{Schwarz2019netspectre}.
Since the latency for a correct guess is in the region of microseconds, not many requests ($\leq 200$) are required per guess to distinguish the latency.
Therefore a simple DDoS detection might detect an attack but only after a certain amount of data being leaked.

\section{Conclusion}\label{sec:conclusion}

In this paper, we presented a timing side channel attack on several memory compression algorithms.
With \Fuzzer, we developed a generic approach to amplify latencies for correct guesses when performing dictionary attack on different compression algorithms.
Our remote covert channel achieves a transmission rate of \SI{643.25}{\bit/\hour} over \SIx{14} hops in the internet.
We demonstrated three cases studies on memory compression attacks.
First, we showed a remote dictionary attack on a PHP application, hosted on an Nginx, storing zlib-compressed data in Memcached.
We demonstrate bytewise leakage across the internet from a server located $14$ hops away. 
Using \Fuzzer, we explored high timing differences when performing a dictionary attack on PostgreSQLs PGLZ algorithm.
and leveraged those to leak database records from PostgreSQL.
Lastly, we also demonstrated a dictionary attack on ZRAM.

\bibliographystyle{plainurl}
\bibliography{main}

\qquad
\FloatBarrier
\appendices
\crefalias{section}{appendix} 
\section{Timing difference for compression}\label{sec:appendix-timing_diff_compression}
\begin{table}[ht]
\setlength{\aboverulesep}{0pt}
\setlength{\belowrulesep}{0pt}
\caption{Different compression algorithms yield distinguishable timing differences when compressing content with a different entropy. ($n=100000$}
\label{tab:evaluated_algorithms_comp}

\begin{center}
\vspace{-0.2cm}	
\resizebox{\hsize}{!}{
    \begin{tabular}{|r|r|r|r|}
        \hline
        \multirow{2}{*}{Algorithm}                    & Fully & Partially & \multirow{2}{*}{Incompressible (ns)} \\
        & Compressible (ns) & Compressible (ns) & \\
        \hline \addlinespace[0.2cm]
        FastLZ & \SIx{38619.07} ($\pm0.74\%$) & \SIx{58887.40} ($\pm0.50\%$) & \SIx{79384.89} ($\pm0.40\%$) \\
        LZ4  & \SIx{606.82} ($\pm0.93\%$) & \SIx{220.63} ($\pm1.00\%$) & \SIx{231.76} ($\pm1.97\%$) \\
        LZ4  & \SIx{44748.02} ($\pm0.15\%$) & \SIx{47731.08} ($\pm0.16\%$) & \SIx{47316.56} ($\pm0.16\%$) \\
        LZO  & \SIx{5645.86} ($\pm2.18\%$) & \SIx{5915.28} ($\pm2.78\%$) & \SIx{7928.21} ($\pm3.91\%$) \\
        PGLZ  & \SIx{44275.84} ($\pm0.13\%$) & \SIx{65752.55} ($\pm0.12\%$) & - \\
        zlib & \SIx{38479.53} ($\pm0.22\%$) & \SIx{80284.72} ($\pm0.23\%$) & \SIx{76973.82} ($\pm0.20\%$) \\
        zstd & \SIx{3596.41} ($\pm0.42\%$) & \SIx{22288.14} ($\pm0.52\%$) & \SIx{29284.77} ($\pm0.34\%$) \\
        \hline
    \end{tabular}
}
    \end{center}
 
\end{table}

\section{Kernel Trace for ZRAM Decompression}\label{sec:appendix-zram-ftrace}
To highlight the root-cause of the timing differences in ZRAM accesses, we trace the kernel functions called on accesses to ZRAM pages using the \texttt{ftrace}~\cite{Rostedt2017ftrace} utility.
\cref{lst:zram_ftrace_rand} and \cref{lst:zram_ftrace_comp} show the trace of kernel functions called on an access to an incompressible and compressible page respectively in ZRAM (using \texttt{deflate} algorithm).
The incompressible page contains random bytes, while the compressible page contains 2048 bytes of the same value and 2048 random bytes similar to the partially-compressible setting in \cref{sec:zram_characterization}.
We only show the functions which are called when the ZRAM page is swapped in to regular memory.

The main difference between the two listings (colored in red) is that the functions performing decompression of the ZRAM page are only called when a compressible page is swapped-in, while these functions are skipped for the page stored uncompressed.
Of these additional function calls, the main driver of the timing difference is the \texttt{\_\_deflate\_decompress} function in \cref{lst:zram_ftrace_comp} which consumes 12555~ns.
This ties in with the characterization study in \cref{sec:zram_characterization} which showed the average timing difference between accesses to compressible and incompressible pages to be close to 10000~ns for ZRAM with \texttt{deflate} algorithm.
These timings are for the \texttt{deflate} implementation in the Linux kernel, which is a modified version of \texttt{zlib v1.1.3}~\cite{Gailly2020zlibh}; hence these timings differ from the \texttt{zlib} timings measured in \cref{tab:evaluated_algorithms_decomp} for the more recent \texttt{zlib v1.2.11}.

\noindent\begin{minipage}{.45\textwidth}
\begin{lstlisting}[caption=Kernel function trace for ZRAM access to incompressible page,frame=tlrb,label={lst:zram_ftrace_rand}]{Incomp}
(*\bfseries Time(ns) Function*)
0          swap_readpage
62         page_swap_info
126        __frontswap_load
195        __page_file_index
254        bdev_read_page
326        blk_queue_enter
395        zram_rw_page
460        zram_bvec_rw.isra.0
527        generic_start_io_acct
590        update_io_ticks
661        part_inc_in_flight
729        __zram_bvec_read.constprop.0
813        zs_map_object
967        _raw_read_lock
1407       zs_unmap_object
1487       generic_end_io_acct
1550       update_io_ticks
1617       jiffies_to_usecs
1677       part_dec_in_flight
1753       ktime_get_with_offset
1835       page_endio
1904       unlock_page
\end{lstlisting}
\end{minipage}\hfill

\begin{minipage}{.45\textwidth}
\begin{lstlisting}[caption=Kernel function trace for ZRAM access to compressible page,frame=tlrb,label={lst:zram_ftrace_comp}]{Comp}
(*\bfseries Time(ns) Function*)
0          swap_readpage
61         page_swap_info
123        __frontswap_load
188        __page_file_index
248        bdev_read_page
310        blk_queue_enter
379        zram_rw_page
442        zram_bvec_rw.isra.0
505        generic_start_io_acct
575        update_io_ticks
634        part_inc_in_flight
755        __zram_bvec_read.constprop.0
838        zs_map_object
1040       _raw_read_lock
(*\aftergroup\speciallstcolor*)1229       zcomp_stream_get(*\aftergroup\endspeciallstcolor*)
(*\aftergroup\speciallstcolor*)1306       zcomp_decompress(*\aftergroup\endspeciallstcolor*)
(*\aftergroup\speciallstcolor*)1373       crypto_decompress(*\aftergroup\endspeciallstcolor*)
(*\aftergroup\speciallstcolor*)1433       deflate_decompress(*\aftergroup\endspeciallstcolor*)
(*\aftergroup\speciallstcolor*)1499       __deflate_decompress(*\aftergroup\endspeciallstcolor*)
(*\aftergroup\speciallstcolor*)14053      zcomp_stream_put(*\aftergroup\endspeciallstcolor*)
14117      zs_unmap_object
14195      generic_end_io_acct
14255      update_io_ticks
14317      jiffies_to_usecs
14374      part_dec_in_flight
14442      ktime_get_with_offset
14518      page_endio
14589      unlock_page
\end{lstlisting}
\end{minipage}

\section{Layout Discovered by \Fuzzer}\label{sec:appendix-zlib-trace}
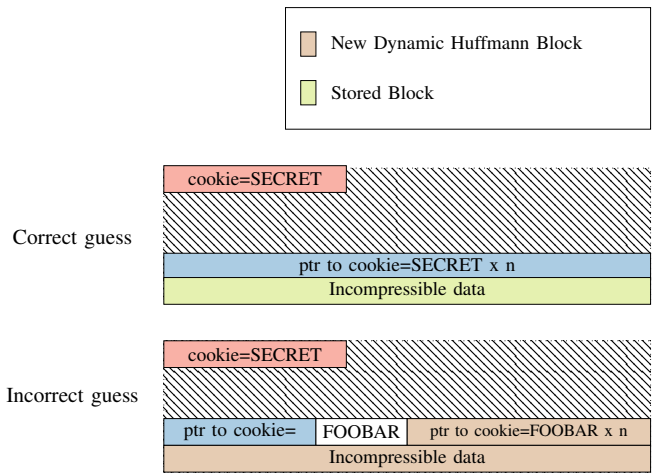
\begin{figure}[ht]
  \centering
  \resizebox{\hsize}{!}{
    \begin{tikzpicture}[yscale=0.8]

\draw[dotted,pattern=north west lines] (-0.5,1) rectangle +(8,2.7);
\draw[dotted,pattern=north west lines] (-0.5,-2.55) rectangle +(8,2.7);

\node at (-2,2.25) {\parbox{2.25cm}{\centering Correct guess}};
\node at (-2,-1) {\parbox{2.25cm}{\centering Incorrect guess}};

\draw[fill=red!40] (-0.5,3.2) rectangle +(3,0.55) node[pos=.5] {\small cookie=SECRET};
\draw[fill=blue!40] (-0.5,1.4) rectangle +(8,0.55) node[pos=.5] {\small ptr to cookie=SECRET x n};
\draw[fill=green!40] (-0.5,0.9) rectangle +(8,0.55) node[pos=.5] {\small Incompressible data};

\draw[fill=red!40] (-0.5,-0.4) rectangle +(3,0.55) node[pos=.5] {\small cookie=SECRET};
\draw[fill=blue!40] (-0.5,-2) rectangle +(2.5,0.55) node[pos=.5] {\small ptr to cookie=};
\draw[fill=yellow!0] (2,-2) rectangle +(1.5,0.55) node[pos=.5] {\small FOOBAR};
\draw[fill=brown!40] (3.5,-2) rectangle +(4.0,0.55) node[pos=.5] {\footnotesize ptr to cookie=FOOBAR x n};
\draw[fill=brown!40] (-0.5,-2.55) rectangle +(8,0.55) node[pos=.5] {\small Incompressible data};

\draw[] (1.5,7) rectangle +(6,-2.5);
\draw[fill=brown!40] (1.75,6) rectangle +(0.25,0.5) node[midway,right,xshift=0.25cm] {\small New Dynamic Huffmann Block};
\draw[fill=green!40] (1.75,5) rectangle +(0.25,0.5) node[midway,right,xshift=0.25cm] {\small Stored Block};
\end{tikzpicture}}

  \caption{Incorrect guesses with the corner-case discovered by \Fuzzer lead to a dynamic Huffman block creation for the partially compressible data, that is slow to decompress.}
  \label{fig:zlib_correct_vs_incorrect}
\end{figure}

\cref{fig:zlib_correct_vs_incorrect} shows the layout discovered by the \Fuzzer that amplifies the timing difference for decompression of the correct and incorrect guesses in Zlib dictionary attack. \cref{lst:zlib_edge_case_trace_correct} and \cref{lst:zlib_edge_case_trace_incorrect} show the debug trace from the Zlib code for decompression with the correct and incorrect guesses to illustrate the root cause of the timing differences.

\begin{minipage}{.45\textwidth}
\begin{lstlisting}[caption=Trace for correct guess in zlib. Here the entire guess string is compressed and the remainder is incompressible (decompressed fast as a stored block).,frame=tlrb, label={lst:zlib_edge_case_trace_correct}]{Comp}
inflate:         length 12
inflate:         distance 16484
inflate:         literal 0x17
inflate:         length 13
inflate:         distance 14
inflate:         literal 0xb3
inflate:         length 13
inflate:         distance 14
inflate:         literal 'x'
inflate:         length 13
inflate:         distance 14
inflate:         literal 0x05
inflate:         length 13
inflate:         distance 14
inflate:         literal 0xa9
inflate:         length 13
inflate:         distance 14
inflate:         literal 0x81
inflate:         length 13
inflate:         distance 14
inflate:         literal '['(*\aftergroup\specialBlstcolor*)
inflate:     stored block (last)
inflate:       stored length 16186
inflate:       stored end(*\aftergroup\endspeciallstcolor*)
inflate:   check matches trailer
inflate: end
\end{lstlisting}
\end{minipage}

\begin{minipage}{.45\textwidth}
\begin{lstlisting}[caption=Trace for incorrect guess in zlib. Here only part of the guess string (\texttt{cookie=}) is compressed and the remainder \texttt{cookie=FOOBAR} is separately compressed (decompression requires a slower code block for Huffman tree decoding). ,frame=tlrb,label={lst:zlib_edge_case_trace_incorrect}]{Comp}
inflate:         length 6
inflate:         distance 16484
inflate:         literal 'F'
inflate:         literal 'O'
inflate:         literal 'O'
inflate:         literal 'B'
inflate:         literal 'A'
inflate:         literal 'R'
inflate:         literal 0x17
inflate:         length 13
inflate:         distance 14
inflate:         literal 0xb3
inflate:         length 13
inflate:         distance 14
inflate:         literal 'x'
inflate:         length 13
inflate:         distance 14
inflate:         literal 0x05(*\aftergroup\specialBlstcolor*)
inflate:     dynamic codes block (last)
inflate:       table sizes ok
inflate:       code lengths ok
inflate:       codes ok
inflate:         length 13
inflate:         distance 14
inflate:         literal 0xa9
inflate:         length 13
inflate:         distance 14
inflate:         literal 0x81
inflate:         length 13
inflate:         distance 14
inflate:         literal '['
inflate:         length 13
inflate:         distance 14
inflate:         literal 0xff
inflate:         length 13
inflate:         distance 14
inflate:         literal 0xa5
inflate:         length 13
inflate:         distance 14(*\aftergroup\endspeciallstcolor*)
\end{lstlisting}
\end{minipage}

\section{Bytewise leakage}\label{sec:appendix-bytewise-leakage}
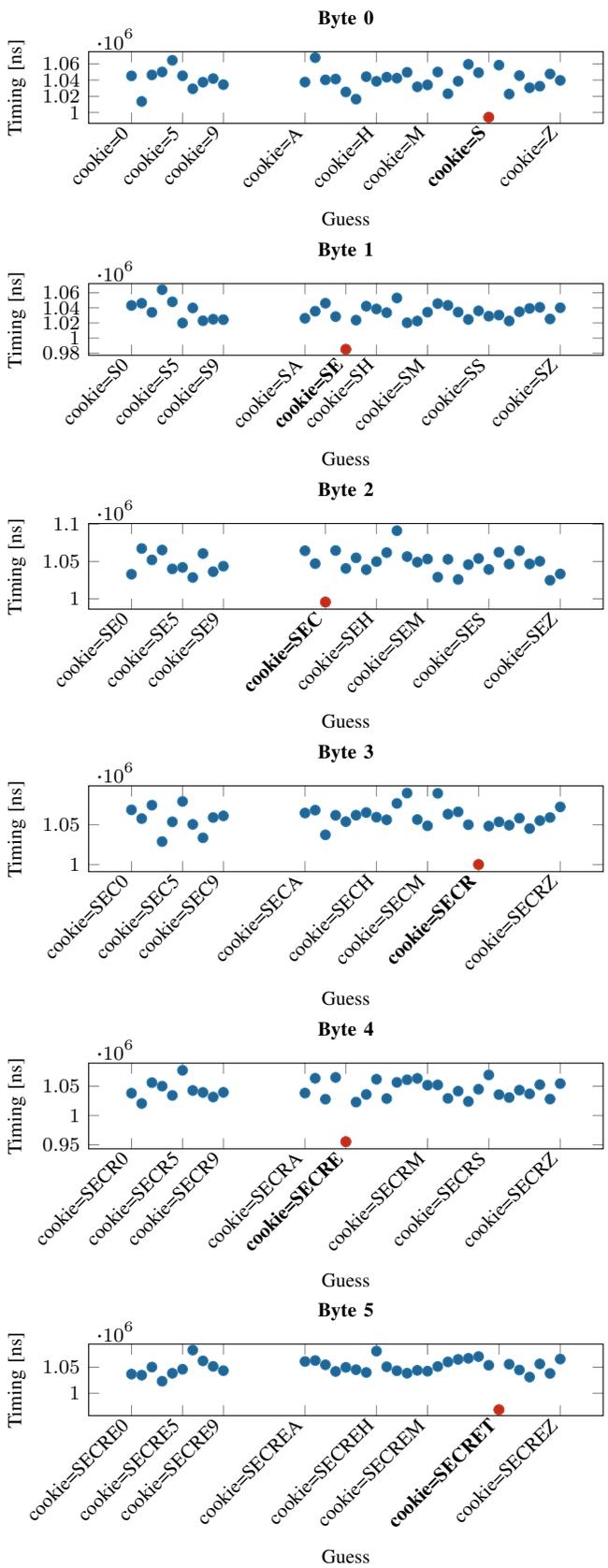
\begin{figure}[t]
 \centering
    \begin{subfigure}{\hsize}%
        \begin{tikzpicture}
\begin{axis}[
title={\textbf{Byte 0}},
style={font=\footnotesize},
xlabel={Guess},
ylabel={Timing [ns]},
y label style={align=center,text width=2cm},
width=\hsize,
height=2.6cm,
xtick={48,53,57,65,72,77,83,90},
xticklabels={cookie=0,cookie=5,cookie=9,cookie=A,cookie=H,cookie=M,\textbf{cookie=S},cookie=Z},
x tick label style={rotate=45,anchor=east}
]

\addplot+[mark=*,draw=none] table[x=Secret,y=Prev_TS_Frame,col sep=comma] {data/memcached_bytewise.csv};
\addplot+[mark=*,only marks,error bars/.cd,y dir=both,y explicit]
        table[x=Secret,y=Prev_TS_Frame,col sep=comma,
        restrict expr to domain={\thisrow{Secret}}{83:83}
        ]{data/memcached_bytewise.csv}; 
\end{axis}
\end{tikzpicture}%
    \end{subfigure}
    \begin{subfigure}{\hsize}%
        \begin{tikzpicture}
\begin{axis}[
title={\textbf{Byte 1}},
style={font=\footnotesize},
xlabel={Guess},
ylabel={Timing [ns]},
y label style={align=center,text width=2cm},
width=\hsize,
height=2.6cm,
xtick={48,53,57,65,69,72,77,83,90},
xticklabels={cookie=S0,cookie=S5,cookie=S9,cookie=SA,\textbf{cookie=SE},cookie=SH,cookie=SM,cookie=SS,cookie=SZ},
x tick label style={rotate=45,anchor=east}
]

\addplot+[mark=*,draw=none] table[x=Secret,y=Prev_TS_Frame,col sep=comma] {data/memcached_bytewise_1.csv};
\addplot+[mark=*,only marks,error bars/.cd,y dir=both,y explicit]
        table[x=Secret,y=Prev_TS_Frame,col sep=comma,
        restrict expr to domain={\thisrow{Secret}}{69:69}
        ]{data/memcached_bytewise_1.csv}; 
\end{axis}
\end{tikzpicture}%
    \end{subfigure}
    \begin{subfigure}{\hsize}%
        \begin{tikzpicture}
\begin{axis}[
title={\textbf{Byte 2}},
style={font=\footnotesize},
xlabel={Guess},
ylabel={Timing [ns]},
y label style={align=center,text width=2cm},
width=\hsize,
height=2.8cm,
xtick={48,53,57,67,72,77,83,90},
xticklabels={cookie=SE0,cookie=SE5,cookie=SE9,\textbf{cookie=SEC},cookie=SEH,cookie=SEM,cookie=SES,cookie=SEZ},
x tick label style={rotate=45,anchor=east}
]

\addplot+[mark=*,draw=none] table[x=Secret,y=Prev_TS_Frame,col sep=comma] {data/memcached_bytewise_2.csv};
\addplot+[mark=*,only marks,error bars/.cd,y dir=both,y explicit]
        table[x=Secret,y=Prev_TS_Frame,col sep=comma,
        restrict expr to domain={\thisrow{Secret}}{67:67}
        ]{data/memcached_bytewise_2.csv}; 
\end{axis}
\end{tikzpicture}%
    \end{subfigure}
    \begin{subfigure}{\hsize}%
        \begin{tikzpicture}
\begin{axis}[
title={\textbf{Byte 3}},
style={font=\footnotesize},
xlabel={Guess},
ylabel={Timing [ns]},
y label style={align=center,text width=2cm},
width=\hsize,
height=2.8cm,
xtick={48,53,57,65,72,77,82,90},
xticklabels={cookie=SEC0,cookie=SEC5,cookie=SEC9,cookie=SECA,cookie=SECH,cookie=SECM,\textbf{cookie=SECR},cookie=SECRZ},
x tick label style={rotate=45,anchor=east}
]

\addplot+[mark=*,draw=none] table[x=Secret,y=Prev_TS_Frame,col sep=comma] {data/memcached_bytewise_3.csv};
\addplot+[mark=*,only marks,error bars/.cd,y dir=both,y explicit]
        table[x=Secret,y=Prev_TS_Frame,col sep=comma,
        restrict expr to domain={\thisrow{Secret}}{82:82}
        ]{data/memcached_bytewise_3.csv}; 
\end{axis}
\end{tikzpicture}%
    \end{subfigure}
    \begin{subfigure}{\hsize}%
        \begin{tikzpicture}
\begin{axis}[
title={\textbf{Byte 4}},
style={font=\footnotesize},
xlabel={Guess},
ylabel={Timing [ns]},
y label style={align=center,text width=2cm},
width=\hsize,
height=2.8cm,
xtick={48,53,57,65,69,77,83,90},
xticklabels={cookie=SECR0,cookie=SECR5,cookie=SECR9,cookie=SECRA,\textbf{cookie=SECRE},cookie=SECRM,cookie=SECRS,cookie=SECRZ},
x tick label style={rotate=45,anchor=east}
]

\addplot+[mark=*,draw=none] table[x=Secret,y=Prev_TS_Frame,col sep=comma] {data/memcached_bytewise_4.csv};
\addplot+[mark=*,only marks,error bars/.cd,y dir=both,y explicit]
        table[x=Secret,y=Prev_TS_Frame,col sep=comma,
        restrict expr to domain={\thisrow{Secret}}{69:69}
        ]{data/memcached_bytewise_4.csv}; 
\end{axis}
\end{tikzpicture}%
    \end{subfigure}
    \begin{subfigure}{\hsize}%
        \begin{tikzpicture}
\begin{axis}[
title={\textbf{Byte 5}},
style={font=\footnotesize},
xlabel={Guess},
ylabel={Timing [ns]},
y label style={align=center,text width=2cm},
width=\hsize,
height=2.6cm,
xtick={48,53,57,65,72,77,84,90},
xticklabels={cookie=SECRE0,cookie=SECRE5,cookie=SECRE9,cookie=SECREA,cookie=SECREH,cookie=SECREM,\textbf{cookie=SECRET},cookie=SECREZ},
x tick label style={rotate=45,anchor=east}
]

\addplot+[mark=*,draw=none] table[x=Secret,y=Prev_TS_Frame,col sep=comma] {data/memcached_bytewise_5.csv};
\addplot+[mark=*,only marks,error bars/.cd,y dir=both,y explicit]
        table[x=Secret,y=Prev_TS_Frame,col sep=comma,
        restrict expr to domain={\thisrow{Secret}}{84:84}
        ]{data/memcached_bytewise_5.csv}; 
\end{axis}
\end{tikzpicture}%
    \end{subfigure}
\caption{Bytewise leakage of the secret (S,E,C,R,E,T) from PHP-Memcached. In each plot, the lowest timing (shown in red) indicates the correct guess.}
\label{fig:memcached_all_bytes}
\end{figure}

\begin{figure}[ht]
  \centering
    \begin{subfigure}{\hsize}%
        \begin{tikzpicture}
\begin{axis}[
title={\textbf{Byte 4}},
style={font=\footnotesize},
ylabel={Timing [ns]},
y label style={align=center,text width=2cm},
width=\hsize,
height=2.8cm,
xtick={48,53,57,65,69,77,83,90},
xticklabels={cookie=SECR0,cookie=SECR5,cookie=SECR9,cookie=SECRA,\textbf{cookie=SECRE},cookie=SECRM,cookie=SECRS,cookie=SECRZ},
x tick label style={rotate=45,anchor=east}
]

\addplot+[mark=*,draw=none] table[x=Char,y=Latency_ns,col sep=comma] {data/zram_bytewise_4.csv};
\addplot+[mark=*,only marks,error bars/.cd,y dir=both,y explicit]
        table[x=Char,y=Latency_ns,col sep=comma,
        restrict expr to domain={\thisrow{Char}}{69:69}
        ]{data/zram_bytewise_4.csv}; 
\end{axis}
\end{tikzpicture}%
    \end{subfigure}
    \begin{subfigure}{\hsize}%
        \begin{tikzpicture}
\begin{axis}[
title={\textbf{Byte 5}},
style={font=\footnotesize},
xlabel={Guess},
ylabel={Timing [ns]},
y label style={align=center,text width=2cm},
width=\hsize,
height=2.6cm,
xtick={48,53,57,65,72,77,84,90},
xticklabels={cookie=SECRE0,cookie=SECRE5,cookie=SECRE9,cookie=SECREA,cookie=SECREH,cookie=SECREM,\textbf{cookie=SECRET},cookie=SECREZ},
x tick label style={rotate=45,anchor=east}
]

\addplot+[mark=*,draw=none] table[x=Char,y=Latency_ns,col sep=comma] {data/zram_bytewise_5.csv};
\addplot+[mark=*,only marks,error bars/.cd,y dir=both,y explicit]
        table[x=Char,y=Latency_ns,col sep=comma,
        restrict expr to domain={\thisrow{Char}}{84:84}
        ]{data/zram_bytewise_5.csv}; 
\end{axis}
\end{tikzpicture}%
    \end{subfigure}

  \caption{Bytewise leakage of the last two bytes of the secret (E,T) from ZRAM. Times for guesses (0-9, A-Z) for each of the bytes are shown. The highest value in each plot  (shown in red)  indicates the correct secret value for the byte.}
  \label{fig:zram_attack_bytewise_byte56}
\end{figure}
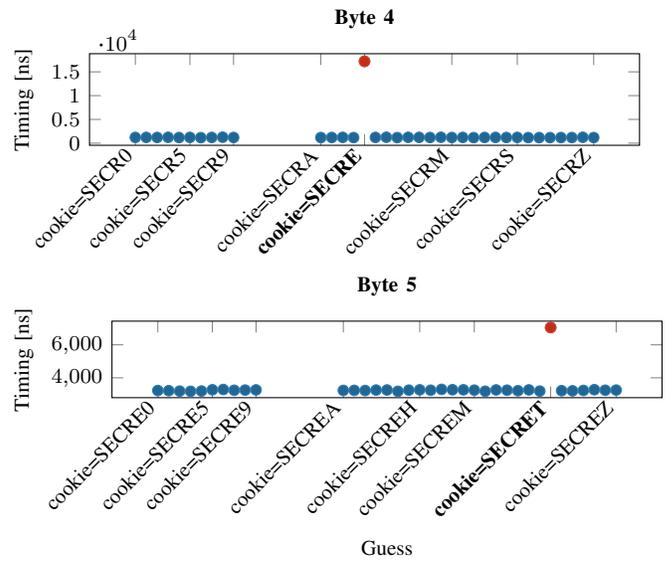

\Cref{fig:memcached_all_bytes} illustrates the bytewise leakage of the secret (SECRET) for a PHP application using PHP-Memcached.
\Cref{fig:postgres_all_bytes} shows the bytewise of the secret string for a Flask application that stores secret data together with attacker-controlled data into PostgreSQL.
The prefix value can be shifted bytewise, which allows reusing the same memory layouts found by \Fuzzer.
\Cref{fig:zram_attack_bytewise_byte56} shows the last two bytes leaked from a secret (SECRET) in a ZRAM page. This is a continuation of \Cref{fig:zram_attack_bytewise_Byte1234} which showed the leakage of the first four bytes.

\begin{figure}[t]
 \centering
    \begin{subfigure}{\hsize}%
        \begin{tikzpicture}
\begin{axis}[
title={\textbf{Byte 0}},
style={font=\footnotesize},
xlabel={Guess},
ylabel={Timing [ns]},
y label style={align=center,text width=2cm},
width=\hsize,
height=2.8cm,
xtick={48,53,57,65,72,77,83,90},
xticklabels={cookie=0,cookie=5,cookie=9,cookie=A,cookie=H,cookie=M,\textbf{cookie=S},cookie=Z},
x tick label style={rotate=45,anchor=east}
]

\addplot+[mark=*,draw=none] table[x=Secret,y=Prev_TS_Frame,col sep=comma] {data/postgres_bytewise_0.csv};
\addplot+[mark=*,only marks,error bars/.cd,y dir=both,y explicit]
        table[x=Secret,y=Prev_TS_Frame,col sep=comma,
        restrict expr to domain={\thisrow{Secret}}{83:83}
        ]{data/postgres_bytewise_0.csv}; 
\end{axis}
\end{tikzpicture}%
    \end{subfigure}
    \begin{subfigure}{\hsize}%
        \begin{tikzpicture}
\begin{axis}[
title={\textbf{Byte 1}},
style={font=\footnotesize},
xlabel={Guess},
ylabel={Timing [ns]},
y label style={align=center,text width=2cm},
width=\hsize,
height=2.6cm,
xtick={48,53,57,65,69,72,77,83,90},
xticklabels={ookie=S0,ookie=S5,ookie=S9,ookie=SA,\textbf{ookie=SE},ookie=SH,ookie=SM,ookie=SS,ookie=SZ},
x tick label style={rotate=45,anchor=east}
]

\addplot+[mark=*,draw=none] table[x=Secret,y=Prev_TS_Frame,col sep=comma] {data/postgres_bytewise_1.csv};
\addplot+[mark=*,only marks,error bars/.cd,y dir=both,y explicit]
        table[x=Secret,y=Prev_TS_Frame,col sep=comma,
        restrict expr to domain={\thisrow{Secret}}{69:69}
        ]{data/postgres_bytewise_1.csv}; 
\end{axis}
\end{tikzpicture}%
    \end{subfigure}
    \begin{subfigure}{\hsize}%
        \begin{tikzpicture}
\begin{axis}[
title={\textbf{Byte 2}},
style={font=\footnotesize},
xlabel={Guess},
ylabel={Timing [ns]},
y label style={align=center,text width=2cm},
width=\hsize,
height=2.6cm,
xtick={48,53,57,67,72,77,83,90},
xticklabels={okie=SE0,okie=SE5,okie=SE9,\textbf{okie=SEC},okie=SEH,okie=SEM,okie=SES,okie=SEZ},
x tick label style={rotate=45,anchor=east}
]

\addplot+[mark=*,draw=none] table[x=Secret,y=Prev_TS_Frame,col sep=comma] {data/postgres_bytewise_2.csv};
\addplot+[mark=*,only marks,error bars/.cd,y dir=both,y explicit]
        table[x=Secret,y=Prev_TS_Frame,col sep=comma,
        restrict expr to domain={\thisrow{Secret}}{67:67}
        ]{data/postgres_bytewise_2.csv}; 
\end{axis}
\end{tikzpicture}%
    \end{subfigure}
    \begin{subfigure}{\hsize}%
        \begin{tikzpicture}
\begin{axis}[
title={\textbf{Byte 3}},
style={font=\footnotesize},
xlabel={Guess},
ylabel={Timing [ns]},
y label style={align=center,text width=2cm},
width=\hsize,
height=2.6cm,
xtick={48,53,57,65,72,77,82,90},
xticklabels={kie=SEC0,kie=SEC5,kie=SEC9,kie=SECA,kie=SECH,kie=SECM,\textbf{kie=SECR},kie=SECRZ},
x tick label style={rotate=45,anchor=east}
]

\addplot+[mark=*,draw=none] table[x=Secret,y=Prev_TS_Frame,col sep=comma] {data/postgres_bytewise_3.csv};
\addplot+[mark=*,only marks,error bars/.cd,y dir=both,y explicit]
        table[x=Secret,y=Prev_TS_Frame,col sep=comma,
        restrict expr to domain={\thisrow{Secret}}{82:82}
        ]{data/postgres_bytewise_3.csv}; 
\end{axis}
\end{tikzpicture}%
    \end{subfigure}
    \begin{subfigure}{\hsize}%
        \begin{tikzpicture}
\begin{axis}[
title={\textbf{Byte 4}},
style={font=\footnotesize},
xlabel={Guess},
ylabel={Timing [ns]},
y label style={align=center,text width=2cm},
width=\hsize,
height=2.6cm,
xtick={48,53,57,65,69,77,83,90},
xticklabels={ie=SECR0,ie=SECR5,ie=SECR9,ie=SECRA,\textbf{ie=SECRE},ie=SECRM,ie=SECRS,ie=SECRZ},
x tick label style={rotate=45,anchor=east}
]

\addplot+[mark=*,draw=none] table[x=Secret,y=Prev_TS_Frame,col sep=comma] {data/postgres_bytewise_4.csv};
\addplot+[mark=*,only marks,error bars/.cd,y dir=both,y explicit]
        table[x=Secret,y=Prev_TS_Frame,col sep=comma,
        restrict expr to domain={\thisrow{Secret}}{69:69}
        ]{data/postgres_bytewise_4.csv}; 
\end{axis}
\end{tikzpicture}%
    \end{subfigure}
    \begin{subfigure}{\hsize}%
        \begin{tikzpicture}
\begin{axis}[
title={\textbf{Byte 5}},
style={font=\footnotesize},
xlabel={Guess},
ylabel={Timing [ns]},
y label style={align=center,text width=2cm},
width=\hsize,
height=2.6cm,
xtick={48,53,57,65,72,77,84,90},
xticklabels={e=SECRE0,e=SECRE5,e=SECRE9,e=SECREA,e=SECREH,e=SECREM,\textbf{e=SECRET},e=SECREZ},
x tick label style={rotate=45,anchor=east}
]

\addplot+[mark=*,draw=none] table[x=Secret,y=Prev_TS_Frame,col sep=comma] {data/postgres_bytewise_5.csv};
\addplot+[mark=*,only marks,error bars/.cd,y dir=both,y explicit]
        table[x=Secret,y=Prev_TS_Frame,col sep=comma,
        restrict expr to domain={\thisrow{Secret}}{84:84}
        ]{data/postgres_bytewise_5.csv}; 
\end{axis}
\end{tikzpicture}%
    \end{subfigure}
\caption{Bytewise leakage of the secret (S,E,C,R,E,T) from PostgreSQL. The known prefix (cookie=) is shifted left by 1 character in each step, allowing the same memory layout to be reused. In each plot, the highest timing (shown in red) indicates the correct guess.}
\label{fig:postgres_all_bytes}
\end{figure}
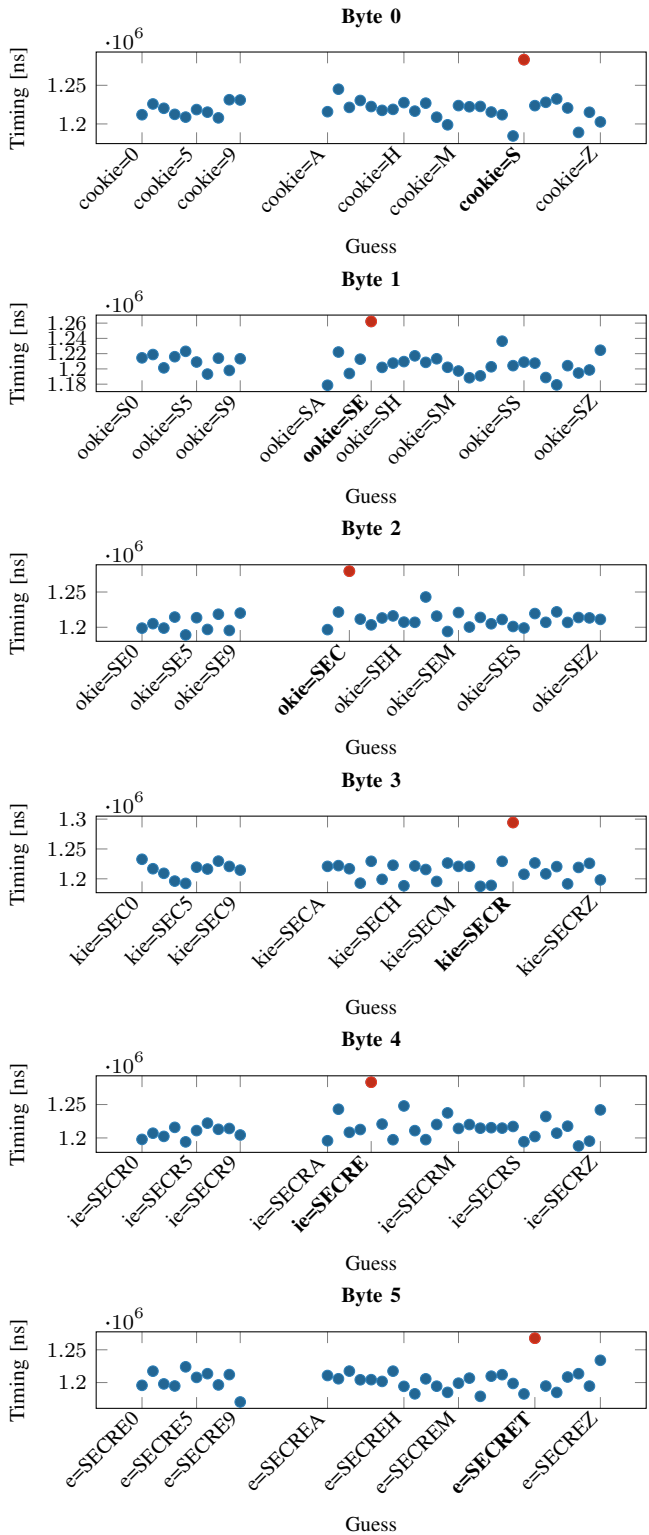

\end{document}